\documentclass{Scipost}
\usepackage[utf8]{inputenc}
\usepackage{hyperref}
\usepackage{bm}
\usepackage{braket}
\usepackage{color}
\usepackage{multirow}
\usepackage{subcaption}
\usepackage[normalem]{ulem} 
\usepackage{float}
\usepackage{graphicx}
\graphicspath{{Figures/}}
\usepackage{mwe} 
\usepackage{amsfonts}
\usepackage{tikz}
\usepackage{rotating}
\usepackage{verbatim}
\usepackage{tablefootnote}
\usetikzlibrary{arrows}
\usetikzlibrary{decorations.pathreplacing,decorations.markings}
\usetikzlibrary{decorations.pathmorphing, patterns,shapes}
\newcommand{\tetra}{
  \mathchoice
    {\includegraphics[height=2ex]{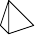}} 
    {\includegraphics[height=2ex]{fig/tetrahedron}} 
    {\includegraphics[height=1.6ex]{fig/tetrahedron}} 
    {\includegraphics[height=1ex]{fig/tetrahedron}} 
}

\newcommand{\sfigref}[2]{Fig.~\hyperref[#1]{\ref{#1}#2}}

\newcommand{\be}{\begin{equation}}
\newcommand{\ee}{\end{equation}}

\newcommand{\ii}{\mathrm{i}}

\begin{document}

\begin{center}{\Large \textbf{
Bulk Excitations of Invertible Phases
}}\end{center}

\begin{center}
Wenjie Ji\textsuperscript{1,4,5},
David T. Stephen\textsuperscript{1,2,$\dagger$},
Michael Levin\textsuperscript{3},
Xie Chen,\textsuperscript{1}
\end{center}

\begin{center}
{\bf 1} Institute for Quantum Information and Matter and Department of Physics, \mbox{California Institute of Technology, Pasadena, CA, 91125, USA}\\
{\bf 2} {Department of Physics and Center for Theory of Quantum Matter, {University of Colorado Boulder}, Boulder, Colorado 80309 USA}\\
{\bf 3} Kadanoff Center for Theoretical Physics, University of Chicago, \mbox{Chicago, Illinois 60637, USA}\\
{\bf 4} Department of Physics and Astronomy, McMaster University, \mbox{Hamilton, Ontario L8S 4M1, Canada}\\
{\bf 5} Perimeter Institute for Theoretical Physics, Waterloo, Ontario, N2L 2Y5, Canada
\\[\baselineskip]
$\dagger$ Current address: {Quantinuum, 303 S Technology Ct, Broomfield, CO 80021, USA}
\end{center}

\begin{center}
\today
\end{center}

\section*{Abstract}
{Recent developments in the study of topological defects highlight the importance of understanding the multi-dimensional structure of bulk excitations inside a quantum system. When the bulk ground state is trivial, i.e. a product state, excitations on top of it are decoupled from each other and correspond to lower-dimensional phases and their defects within. In this paper, we expand the discussion to invertible phases and study the bulk excitations in, for example, SPT phases, majorana chains, $p+ip$ superconductors etc. We find that there is a one-to-one correspondence between bulk excitations inside a nontrivial invertible phase and those in a product state. For SPT phases, this can be shown using the symmetric Quantum Cellular Automaton that maps from the product state to the SPT state.  More generally, for invertible phases realizable using the Topological Holography construction, we demonstrate the correspondence using the fact that certain gapped boundary conditions of a topological bulk state have only relative distinctions but no absolute ones.
}

\tableofcontents

\section{Introduction}

In noninteracting systems, particle excitations carry the full information about the bulk spectrum. In interacting systems, sometimes a quasi-particle picture also works well (Fermi-liquids, fractional Quantum Hall states, for example). But there are also cases where higher-dimensional excitations are of fundamental importance. Examples include the Ising ferromagnetic phase in $2+1$D and discrete gauge theories in $3+1$D. In the ferromagnet, the basic form of excitation is the magnetic domain walls, which form loops. Flux loops in gauge theories, on the other hand, underlie the ground state degeneracy on nontrivial manifolds and have nontrivial braiding statistics with the point charge excitation. To properly understand the excitation spectrum in these systems and other strongly interacting systems in general, it is important to consider higher-dimensional excitations as well.

Let us start from the simplest case: a trivially gapped $D$-dimensional bulk state.\footnote{In the following discussion, we use $D$ to denote the dimension of the bulk system and use $d$ to denote the dimension of the excitations. $d<D$.} This can be an atomic insulator or a polarized spin state whose ground state wave function can be continuously connected to a product state. Usually we think about point excitations above such a state -- electrons in a higher orbital in the atomic insulator or a spin flip in the polarized spin state -- and the waves formed by such point excitations. But there are also line excitations and even surface excitations. In a product-state bulk, such $d$-dimensional excitations are unentangled from the bulk and from other excitations, and they form their own state. The collection of $d$-dimensional excitations in a product state bulk hence has the same structure as $d$ dimensional quantum systems. But $d$-dimensional excitations can come in different types and shapes. What are the interesting ones that we want to consider and how to formulate their structure? For the product-state bulk, since the $d$-dimensional excitations correspond to $d$-dimensional quantum states, one choice is to focus on those that correspond to the ground states of $d$-dimensional gapped phases. One way to understand gapped phases is with quantum circuits \cite{Chen2010}. This picture can be summarized as follows: \textit{States within the same gapped phase are connected through finite depth circuits while states in different phases are connected through sequential quantum circuits.}\footnote{We will always allow the freedom to add a finite density of ancilla degrees of freedom in the form of product states in these equivalence relations.} A sequential quantum circuit is a kind of circuit that preserves the entanglement area law \cite{chen2024sequential}. According to this definition, gapped states generated from a product state with a finite depth circuit belong to the trivial phase. With this equivalence relation, gapped quantum states fall into discrete gapped phases and the study of the structure of $d$-dimensional excitations above the product-state bulk now becomes the study of the structure of $d$-dimensional gapped phases.

For non-product bulk states, excitations within are coupled to the bulk and hence have different structures. A similar criterion can be set up to study $d$-dimensional excitations above a bulk state in general:
\begin{enumerate}
\item We will focus on $d$-dimensional excitations that are gapped ground states of a modified Hamiltonian with Hamiltonian terms differing from the bulk only along the excitation.
\item If two $d$-dimensional excitations can be mapped into each other through a $d$-dimensional finite-depth circuit, then the two are equivalent excitations.
\item $d$-dimensional excitations generated from the bulk ground state with a $d$-dimensional finite-depth circuit are of the trivial type. 
\item Nontrivial types of $d$-dimensional excitation cannot be generated from the bulk with a $d$-dimensional finite-depth quantum circuit. They are generated with either a $d+1$-dimensional circuit or a $d$-dimensional sequential circuit. 
\end{enumerate}

This set of criteria is a natural generalization of the notion of `super-selection sector' for quasi-particle excitations. Equivalence relations between quasi-particle excitations are given by $0$-dimensional finite depth circuits -- that is, local unitary transformations. Nontrivial types of quasiparticle excitations have to be generated with a `string operator -- a $1$-dimensional finite-depth circuit or sequential circuit. While we do not usually think of quasi-particles as ground states of a modified Hamiltonian, we can. We can modify the Hamiltonian so that it traps a quasi-particle excitation at a particular location. In fact, when we talk about the `degeneracy' of nonabelian anyons, it is with respect to this modified Hamiltonian not the original bulk Hamiltonian. In higher dimensions, this set of criteria covers the situations mentioned above: domain walls in $2+1$D ferromagnets and flux loops in $3+1$D gauge theories. 

\begin{figure}[ht]
    \centering
    \includegraphics[scale = 0.5]{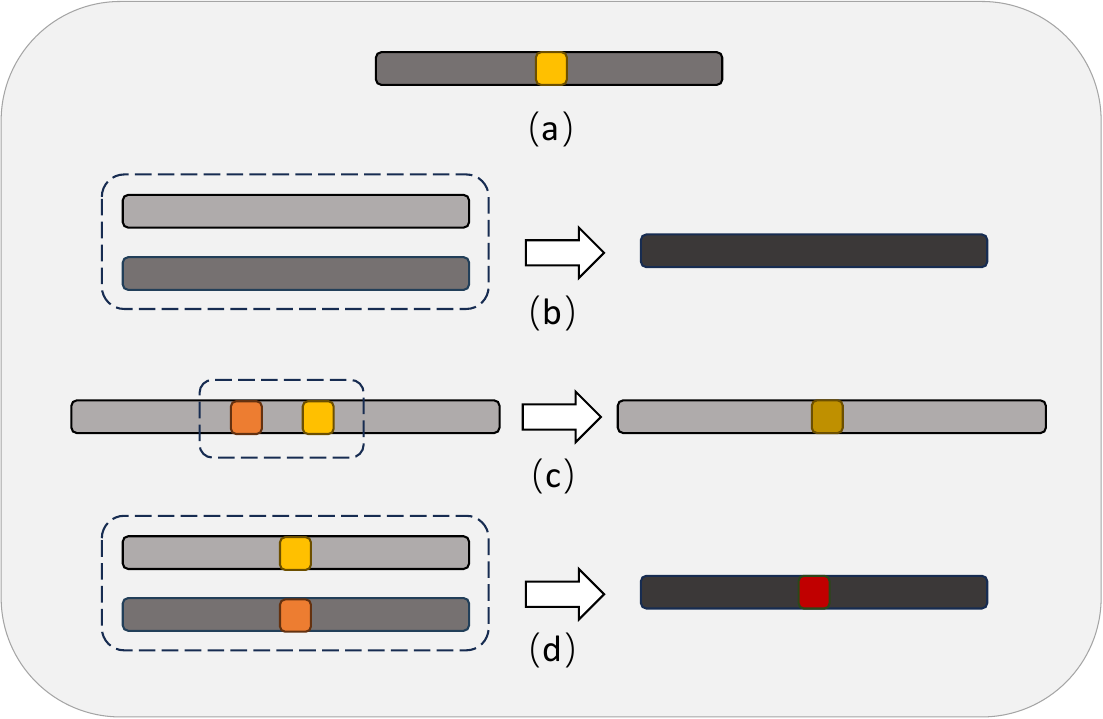}   
    \caption{Multi-dimensional excitations and their fusion. (a) a 1-dimensional excitaiton with a 0-dimensional excitation on top. (b) fusion of two 1-dimensional excitations using a 1-dimensional finite depth circuit. (c) fusion of two 0-dimensional excitations on top of a 1-dimensional excitation using a 0-dimensional local unitary operation. (d) fusion of two 1-dimensional excitations with 0-dimensional excitations on top using a 1-dimensional finite depth circuit.}
    \label{fig:fusion}
\end{figure}

We can use the above criterion recursively to establish a `higher-category' structure of excitations which highlights the fact that not only do excitations come in different dimensions, but excitations from different dimensions are inherently interconnected with each other\cite{thorngren2019fusion,Johnson-Freyd2022,Kong2020,Bhardwaj2023}. As shown in Figure~\ref{fig:fusion}, on top of the $d$-dimensional excitations, we can consider $d-1$-dimensional excitations which are the gapped ground state of a Hamiltonian modified further along a $d-1$-dimensional submanifold within the $d$-dimensional excitation. The equivalence relation among the $d-1$-dimensional excitations are given by $d-1$-dimensional finite depth circuits.\footnote{$d-1$-dimensional excitations that stand alone in the bulk can be thought of as on top of a trivial $d$-dimensional excitation.} Then on top of the $d-1$ dimensional excitations (on top of a $d$-dimensional excitation), we can further consider $d-2$-dimensional excitations with equivalence relation given by $d-2$ dimensional finite depth circuits and so on. In the category language, the $d$-dimensional excitations are called `objects', the $d-1$-dimensional excitations on top are called `morphisms', and the $d-2$-dimensional excitations are called `morphisms of morphisms', etc. Given the set of excitations, we can further talk about the fusion among them. Fusion between a pair of $k$-dimensional excitations is given by a $k$-dimensional finite depth circuit that maps the pair of excitations to their fusion result (another $k$-dimensional excitation). For example, we can study the fusion of two $d$-dimensional excitations, the fusion of two $d-1$-dimensional excitations on top of a $d$-dimensional excitation, and the fusion of two $d$-dimensional excitations with $d-1$-dimensional excitations on top, as shown in Figure~\ref{fig:fusion}. By establishing such a `fusion higher-category' description of bulk excitations, we expect to have a better approach for studying higher-dimensional strongly interacting quantum phases and phase transitions.

When the system under consideration is constrained by a symmetry, a `symmetric' version of the above definitions apply. Both the original Hamiltonian and the modified Hamiltonian are composed of local symmetric terms. The equivalence between $d$-dimensional excitations are given by symmetric $d$-dimensional finite-depth circuits, where each local gate in the circuits is symmetric and the ancillas added are in symmetric product states. Most of the examples considered in this paper belong to systems with symmetries. For example, fermionic $\mathbb{Z}_2$ parity symmetry, bosonic $\mathbb{Z}_2\times \mathbb{Z}_2$ symmetry, etc.

In Ref.~\cite{Stephen2024}, we studied the fusion structure of $1$-dimensional gapped phases, hence the fusion structure of $1$-dimensional excitations inside a higher-dimensional product-state bulk. While the excitation structure of a product-state bulk is relatively easy to identify -- they are given by the $d$-category structure of $d$-dimensional gapped quantum phases -- that of a non-product bulk is not so obvious. For example, in a non-trivial SPT state or the $p+ip$ superconducting state, the ground state wave function is always entangled, so it is not immediately obvious what the full excitation structure should be. In this paper, we argue that invertible phases -- phases like SPT and $p+ip$ superconductor which when combined with an inverse become equivalent to the trivial phase -- have the same excitation structure as the trivial product state, which forms the same $d$-category as $d$-dimensional gapped quantum phases. For SPT phases within the group cohomology classification, this equivalence can be shown with a simple argument based on the symmetric Quantum Cellular Automaton that maps from the trivial phase to the nontrivial SPT phase. We present this argument in section~\ref{sec:QCA}. For chiral invertible phases like the $p+ip$ superconductor, the QCA argument does not work. In this case, we establish the result using a `pumping' procedure in one higher dimension, as discussed in section~\ref{sec:p+ip}. Moreover, in section~\ref{sec:symTFT}, we discuss this problem in the Topological Holography context. The Topological Holography framework, also referred to as the Symmetry Topological-Field-Theory or the Symmetry Topological Order framework\cite{KZ150201690,Ji2020,Kong2020,Lichtman2021,Chatterjee2023symmetry,Moradi2022topological,Freed2023topological,lin2023asymptotic,kong2018gapless,kong2021mathematical,kong2022one,kong2022categories,kong2024categories,xu2024categorical}, realizes $D$-dimensional quantum phases using a sandwich structure with a $D+1$-dimensional topological theory in the bulk. Our argument in section~\ref{sec:symTFT} starts by pointing out in section~\ref{sec:relative} that certain gapped boundary conditions of a topological bulk state have only relative differences but no absolute ones. This property is then used in section~\ref{sec:example} to establish the relation between the excitations in invertible phases and those in trivial phases. To describe fermionic systems with fermion parity symmetry within the topological holography framework, fermion condensation \cite{aasen2019fermion} and fermion condensed boundary conditions are necessary. We thus introduce them concretely in lattice models in Appendix \ref{app:fcondensedbdry}. Before getting into these arguments, we discuss in section~\ref{sec:defects} the relation between the notion of defects and what we call $d$-dimensional excitations here. It is not essential for understanding the following sections but may clarify some confusion about our perspective and terminology. 

\section{Defects and Low-Entanglement Excitations}
\label{sec:defects}

The $d$-dimensional excitations we discussed above are usually called `defects' -- changes in the ground state wavefunction induced by the modification of Hamiltonian terms along the defect.\footnote{A symmetry defect is a prototypical example where the change in Hamiltonian is induced by applying symmetry to a sub-region in the system. Hamiltonian terms on the boundary of the sub-region changes by conjugating half of each term with the symmetry. Hamiltonian terms away from the boundary remain invariant.} With this setup, we can talk about the defect being gapped or gapless if the wavefunction with defect is the gapped or gapless (with gapless excitations along the defect) ground state of the modified Hamiltonian. As the ground state of the modified Hamiltonian, defects are static and not dynamical. 

Instead, in this paper, we want to take the point of view that these $d$-dimensional objects are excitations of the original Hamiltonian. We usually do not call such objects with $d>0$ `excitations' because they are usually not eigenstates of the original Hamiltonian and, moreover, have extensive energy. Usually we are interested in low-energy excitations that contribute to response functions. That worked well for point excitations because their energy is at most finite. Excitations with $d>0$ have extensive energy and are high up in the dense part of the spectrum. 

How do we then separate these excitations from the rest of the spectrum and study their structure? Being the ground state of a modified Hamiltonian guarantees that the wave function with the $d$-dimensional excitations still has entanglement area law. Therefore, these excitations have `low entanglement' (instead of low energy). The set of entanglement area law preserving $d$-dimensional excitations is the object of interest here, and we will call them `Low-Entanglement Excitations' (LEE). These low-entanglement excitations will form the higher-category structure discussed above. Being low-entanglement, these excitations can be studied starting from the ground state wavefunction without reference to the original or modified parent Hamiltonians. They are modifications of the ground state wavefunction that preserves the entanglement area-law and their equivalence is given by finite depth circuits. Therefore, we expect that the structure of the LEEs to be the same for the same ground state wave-function with different parent Hamiltonians. Their exact energy can be different with respect to different parent Hamiltonians, but their higher category structures remain the same.

As excitations, we can talk about their dynamics and stability. Since these LEEs are in general not eigenstates, they will deform and change under the dynamics induced by the Hamiltonian. As the Hamiltonian has a higher energy only along the defect, the induced dynamics is concentrated near the $d$-dimensional manifold. Now since the equivalence class of the $d$-dimensional LEEs are given by finite depth $d$-dimensional circuits, the excitations will remain stable (be in the same class) for a finite time under the dynamics induced by the Hamiltonian. On the other hand, we can also design the unitary operations to be applied to move, deform and fuse the $d$-dimensional excitations, all with finite depth circuits \cite{Tantivasadakarn2024}. 

What is the physical importance of LEEs? Since LEEs with $d>1$ have extensive energy, they do not play any role in the linear response of the system under external perturbation. But with low entanglement, the LEEs can potentially be condensed, driving phase transitions between different gapped phases. Once the LEEs are condensed, they do not cost energy any more. Moreover, with low entanglement, their condensate can also have low entanglement (area law), and potentially realize a different gapped phase. Therefore, we expect LEEs to play an important role in the description of phase transitions. 

\section{Symmetric QCA argument}
\label{sec:QCA}

\subsection{General Argument}

The equivalence of the excitation structure of an invertible phase and that of a trivial product state can be easily established for a large classes of invertible phases: those that can be prepared from a trivial product state by a quantum cellular automata (QCA). On a quantum many-body lattice system, a QCA is a locality-preserving unitary operator \cite{Farrelly2020reviewofquantum}. Under conjugation by a QCA, local operators are mapped to local operators, and the operator algebra is preserved. A simple example of a QCA is a finite depth transformation of local unitaries (FDLU) (a.k.a., finite depth quantum circuit), but there are also QCA that are not FDLUs such as the lattice translation operator.

In the presence of a symmetry, we say that a QCA is symmetric if it commutes with all symmetry operators. Hence, a symmetric QCA maps symmetric local operators to symmetric local operators and charged local operators to local operators of the same charge. A symmetric QCA is trivial if it is a symmetric FDLU, \textit{i.e.} an FDLU for which each gate individually commutes with all symmetry operators. A typical example of non-trivial symmetric QCAs in a bosonic lattice system is the SPT entangler, which is an FDLU that commutes as a whole with the symmetry operators but individual gates do not so that the circuit can map a trivial product state to an SPT state \cite{Chen2013}. We want to emphasize that here we define a symmetric QCA to be a QCA that is symmetric. That is, it is locality preserving on all operators and further more symmetric. We want to contrast it with, for example, the Kramers-Wannier transformation\cite{Kramers1941}, which maps symmetric local operators to symmetric local operators but does not preserve the locality of charged operators.

For each SPT phase within the group cohomology classification, we can identify an SPT entangler that generates a fixed-point state from the trivial product state \cite{Chen2013}. In such fixed-point states, we can straightforwardly argue that the classification of LEEs is the same as in the trivial phase. Suppose we start in the trivial phase and insert a $d$-dimensional LEE. Then, apply the SPT entangler to the entire system. This results in a system belonging to the SPT phase with a new defect. Because the SPT entangler is a symmetric QCA, it maps the original LEE onto a new defect with the same dimension and support (up to a constant increase). Furthermore, since applying a QCA preserves the area law, the new defect still has area law entanglement. Therefore, it defines an LEE in the SPT phase. Similarly, we can start with an LEE in the SPT phase and apply the inverse SPT entangler to obtain the corresponding LEE in the trivial phase. Finally, if two LEEs in one phase are related by a symmetric FDLU, then this FDLU can be pulled through the SPT entangler to obtain an operator that relates the two corresponding LEE in the other phase. Since the SPT entangler is a symmetric QCA, it maps symmetric FDLUs onto symmetric FDLUs, so this new operator is again a symmetric FDLU. Therefore, the classifications of LEEs are the same in the trivial and group cohomology SPT phases. If we further assume that all LEEs in the trivial phase can be created using a sequential quantum circuit, as was argued in Ref.~\cite{chen2024sequential}, then the same reasoning as above shows that this is also true for all LEEs in SPT phases.

The above argument can be extended to some SPT phases beyond the group cohomology classification. In particular, Ref.~\cite{Fidkowski2020} constructed a 4-dimensional state that belongs to a beyond cohomology SPT phase. They further showed that this state can be prepared from a trivial product state using a symmetric QCA. Therefore, the above argument applies verbatim to this example. 

\subsection{Example: 1d cluster state}

As a simple example of the symmetric QCA argument, consider the 1D cluster state defined on a lattice of $2N$ qubits with periodic boundary conditions \cite{Briegel2001},
\begin{equation}
    |C\rangle = U_C|++\cdots +\rangle,
\end{equation}
where,
\begin{equation}
    U_C = \prod_{i=1}^{2N} CZ_{i,i+1},
\end{equation}
is the SPT entangler for the cluster state with $CZ = |0\rangle\langle 0|\otimes I + |1\rangle\langle 1|\otimes Z$. The cluster state is also the ground state of the Hamiltonian $H_C = -\sum_i Z_{i-1}X_iZ_{i+1}$ since it satisfies $\langle C|Z_{i-1}X_iZ_{i+1}|C\rangle = 1$. The state possesses non-trivial SPT order with respect to the $Z_2\times Z_2$ symmetry generated by $X_{\mathrm{even}} = \prod_{i=1}^N X_{2i}$ and $X_{\mathrm{odd}} = \prod_{i=1}^N X_{2i-1}$ \cite{Son2011}. This non-trivial SPT order can be seen from the non-trivial string order $\langle C| Z_{i-1} X_i X_{i+2} \cdots X_{i+2L} Z_{i+2L+1}|C\rangle = 1$ which implies, among other things, that the cluster state cannot be prepared from a trivial product state using a symmetric FDLU \cite{huang2015quantum}.

One can confirm that the entangler $U_C$ commutes with both of these symmetries as a whole, but the individual $CZ$ gates do not commute, so $U_C$ is a non-trivial symmetric QCA. In the trivial phase represented by $|++\cdots +\rangle$, the distinct class of non-trivial LEEs can be represented by symmetry charges. These are created by applying $Z$ at a single even or odd site, or a pair of neighbouring sites. In the present case, these $Z$ operators commute with the entangler $U_C$, so they also produce distinct LEEs when acting on the cluster state. One can try to make a new kind of LEE in the cluster state by inserting a symmetry flux, which is accomplished by conjugating all Hamiltonian terms by a symmetry operator acting on half the system. But, this turns out to be equivalent to inserting a symmetry charge. To be concrete, suppose we pick an even site $j$ and conjugate all Hamiltonian terms by the operator $\bar{X}_{\mathrm{even}}= \prod_{i\geq j} X_{2i}$. This has the effect of flipping the sign of the Hamiltonian term at site $i=j-1$, which can also by accomplished by instead inserting a symmetry charge (conjugating with $Z$) at site $j-1$. Therefore, we see that the LEEs in the trivial phase and SPT phase containing the cluster state are identical.

\section{The $p+ip$ Superconductor case}
\label{sec:p+ip}

The QCA argument used in the previous section does not apply to chiral invertible phases like the $p+ip$ superconductor in $2+1$D. {This is because states created from a product state using a QCA necessarily have zero correlation length, wheras chiral invertible phases require a non-zero correlation length.} In this section, we make use of a `pumping' procedure in one higher dimension and argue that the low-entanglement excitations (LEE) in the $2+1$D $p+ip$ superconducting state have the same classification as the LEEs in the trivial $2+1$D superconductor.

\subsection{Pumping $p+ip$ through $3+1$D bulk}

\begin{figure}[ht]
    \centering
    \includegraphics[scale=0.6]{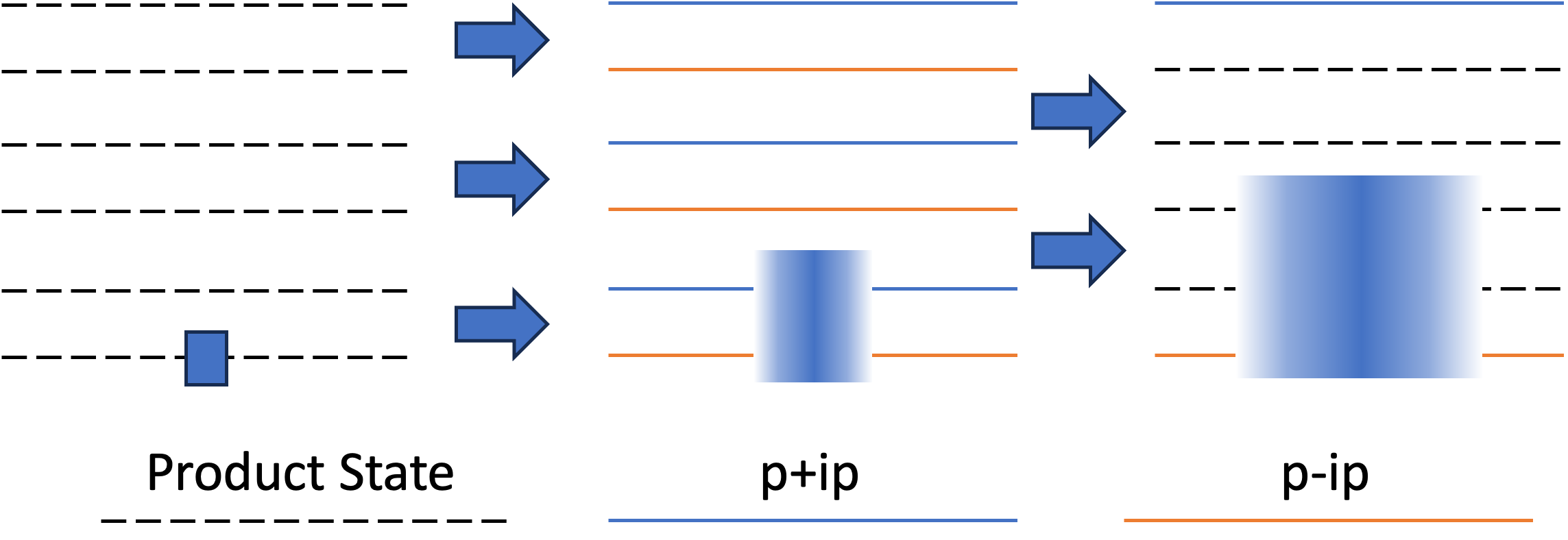}
    \caption{The pumping procedure that generates a $p+ip$ state in the top layer and a $p-ip$ state in the bottom layer of a 3D system while the bulk of the system remains in the trivial product state. The procedure can be realized through a finite-time evolution with a quasi-local Hamiltonian. The blue box indicates a strictly local operator which, after the procedure, is mapped to an operator strictly local in the $z$ direction and quasi-local (with exponentially decaying tails indicated by the color gradient) in the $x$ and $y$ directions.}
    \label{fig:pump}
\end{figure}

First, we describe the process of pumping a $p+ip$ state across multiple layers in a 2D system using a finite time evolution. Consider a 2D layered system with fermion modes on the lattice sites in each layer. Suppose that we start from the trivial product state $|000...0\rangle$ with no fermion on each lattice site. This is the ground state of the Hamiltonian $H = \sum_i c_i^{\dagger}c_i$. Now take every two horizontal $xy$ layers (layer $2k-1$ and $2k$) and adiabatically evolve them to a pair of $p+ip$ and $p-ip$ states. Next, shift one layer and take the pair of $p-ip$ state on layer $2k$ and the $p+ip$ state on layer $2k+1$ and adiabatically evolve them back to the trivial product state. If we apply this procedure to a system with the top layer connected to the bottom layer, then the whole system returns to the product state. If we apply this to a system with open boundaries in the $z$ direction, then in the second step, the evolution does not involve the top and bottom layer and they remain in the $p+ip$ and $p-ip$ states, respectively, while the middle part of the system goes back to the product state. 

This adiabatic process can be converted into a finite-time evolution with a quasi-local Hamiltonian,
\begin{equation}
U = \mathcal{T} \text{exp} \left\{i \int_0^T dt \mathcal{D}_t\right\}
\end{equation}
where $\mathcal{T}$ denotes time ordering and $T$ is a finite constant independent of system size. The Hamiltonian generating the evolution $\mathcal{D}_t$ can be obtained from the quasi-adiabatic continuation procedure\cite{Hastings2005}. Note that because the adiabatic process involves two layers at a time, $U$ is strictly locality preserving in the $z$ direction. But it is quasi-local in the $x$ and $y$ directions. That is, if we conjugate a (strictly) local operator by $U$, it is going to acquire (generically) exponentially decaying tails in the $x$ and $y$ directions but not in the $z$ direction. It can grow in the $z$ direction by at most a distance of four.  

Note that $U$ does not help us to prepare a single $p-ip$ layer since it also creates a corresponding $p+ip$ layer at the top. Even if we truncate $U$ to act only on the Hamiltonian terms of the bottom $m$ layers in the product state, it does not give us a 2D Hamiltonian for the $p-ip$ state. This is because the transformed Hamiltonian terms act on $m+2$ layers such that we need to add more Hamiltonian terms to avoid extensive degeneracy, and the added terms will inevitably generate a $p+ip$ state. Therefore, $U$ does not induce a QCA for preparing a $p-ip$ state in a strictly 2D system, even if we allow for exponentially decaying tails. However, as we show below, it is still sufficient for proving the equivalence of LEE classes between the $p+ip$ and trivial superconductors.

\subsection{Low-entanglement excitations}

Now we can argue for the equivalence in the classification of low-entanglement excitations in the trivial and the $p-ip$ state. If we start from the trivial state (on the left-hand side of Figure~\ref{fig:pump}) with some LEE within the bottom $n$ layers, applying $U$ would map the system to the $p+ip$ state at the top and the $p-ip$ state at the bottom (on the right-hand side of Figure~\ref{fig:pump}) with some LEE in the bottom $n+2$ layers. The layers above the bottom $n+2$ layers are still in the product state and will not be affected by the existence of the LEE (because of the strict locality of $U$ in the $z$-direction). If the LEE before the mapping is gapped (has entanglement area law), the LEE after mapping is gapped (has entanglement are law) as well. 
The change in the number of layers involved can be accommodated by our freedom to add a finite density of ancillas in relating LEEs. Moreover, now suppose we have two such LEEs in the trivial state and they can be connected through a finite depth circuit acting within the bottom $n$ layers. Then applying $U$ we get a finite-depth circuit mapping between the two LEEs in the $p-ip$ state within the bottom $n+2$ layers. Again, the change of number of layers can come from the addition of ancillas. This argument also runs in the reverse direction. Therefore, we see that the LEE equivalence classes for the trivial state and those for the $p-ip$ state are the same. 

A similar argument can be applied to other chiral invertible states, such as the Chern insulator, the $E_8$ state, etc. 

\section{Topological Holography argument}
\label{sec:symTFT}

\begin{figure}[h]
    \centering
    \includegraphics[scale=0.5]{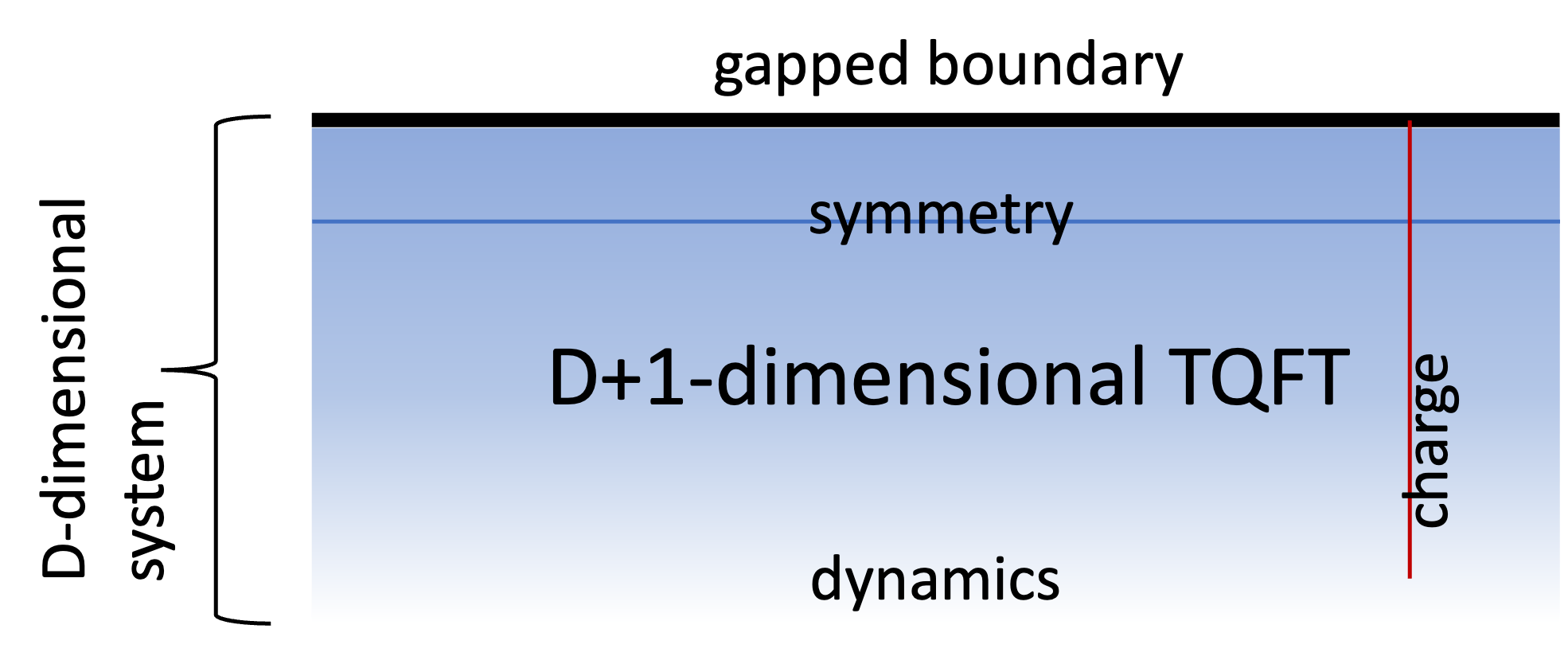}
    \caption{The `sandwich' realization of a $D$-dimensional system in the Topological Holography formalism. In this formalism, a $D$-dimensional system is realized as a ‘sandwich’ structure with a $D+1$-dimensional topological bulk. The bulk and the gapped top boundary determines the symmetry of the $D$-dimensional system, while the bottom boundary contains all its (symmetric) dynamics.}
    \label{fig:sandwich}
\end{figure}

In this section, we establish the equivalence of LEEs in different invertible phases using the Topological Holography framework. The Topological Holography framework realizes a $D$-dimensional system using a ``sandwich'' structure with a $D+1$-dimensional topological theory in the bulk, as shown in Figure~\ref{fig:sandwich}. The top boundary is set to be in a gapped state which, together with the bulk, determines the symmetry of the ``sandwich''. The bottom boundary contains all the dynamical terms. Therefore, the sandwich structure separates the symmetry (the kinetic aspect) of the system from its dynamics and realizes the symmetry in a topological way. 

In the sandwich structure, different invertible phases can be realized by setting the bottom boundary in different gapped states. Due to the invertible nature of the phases, the boundary states can be related to each other through a symmetry transformation of the bulk. The symmetry transformation generated by a FDLU relates any experiments happening near one boundary to those near the other. Therefore, two such boundaries cannot be \textit{absolutely} distinguished. The only way to tell their difference is to put the two boundary side by side and perform experiments across the domain wall between the two. We explain this point in section~\ref{sec:relative} through several examples. This point may be of interest on its own.

Using the relativeness of the distinction between different gapped boundaries, we establish the equivalence of LEEs in the invertible phases realized with the corresponding ``sandwiches''.  This is discussed, again with several examples, in section~\ref{sec:Invert_symTFT}. While the same conclusion was already established in section~\ref{sec:QCA} and section~\ref{sec:p+ip}, the Topological Holography framework can provide a more natural setting to understand the equivalence between trivial and nontrivial invertible phases. One important distinction between the trivial and non-trivial invertible phases is that the trivial phase can have the product state as its ground state while the nontrivial phases do not have such a limit. Such a distinction can be made when we have a tensor product structure of the underlying Hilbert space. When the invertible phases are realized in the ``sandwich'' structure, the tensor product structure of the Hilbert space is automatically lost, as the theory is constrained to live in the low-energy subspace fixed by the bulk and the top boundary Hamiltonian terms. Without a tensor product structure, the realization of the trivial and non-trivial invertible phases in the sandwich can have very similar structures. For example, in some cases, two phases can be related to each other through a bulk on-site unitary $\mathbb{Z}_2$ symmetry. It is then actually impossible to define unambiguously which one is the trivial phase and which one is the non-trivial phase. The only way to see their distinction is by putting the two phases side-by-side and detecting non-trivial edge modes on the domain wall, i.e. only the difference between two phases is well-defined.

\subsection{Relative distinction between gapped boundaries of topological states}
\label{sec:relative}
Let us now explain the relative distinction based on the topological holography construction concretely. Topological ordered phases have emergent symmetries. For examples, in two dimensions, point-like topological excitations can be permuted.  On a spatial manifold without a boundary, the unitary operators generating the symmetry leave the ground state subspace invariant. However, they act non-trivially on the excited states, as well as within the ground state subspace (or rather, the code space). As a result, on a spatial manifold with a boundary, in general, the symmetry is no longer preserved on the boundary. Specifically, the boundary of a topological state can be gapped when topological excitations are condensed on the boundary. Now consider applying the unitary to the whole space. The bulk is still in a ground state. On the boundary, if certain condensed topological excitations are permuted with uncondensed ones, the boundary state is then changed to another gapped one. Thus different boundary states are also permuted by the unitary.

\subsubsection{The Claim}

We claim that two gapped boundary states of a topological order that are related by a finite depth transformation of (quasi) local unitaries (FDLU or FDqLU) that leaves the bulk in a ground state, cannot be distinguished by any local experiments near the boundary. By local experiments, we allow, for example, local unitary transformations, or local measurements, and the results are recorded as classical $\mathbb{C}$-numbers. We refer to two states that cannot be distinguished by local experiments as having no \emph{absolute} distinction.

We further claim that, the two boundary states, though without an absolute difference, can always be distinguished by experiments performed across the boundary between the two. Thus, the two boundary states have \emph{relative} distinctions.

\subsubsection{The Argument}
Now we give an argument for the claim. First we describe a property of FDLUs. The FDLU when acting on an operator by conjugation, maps a local operator to a local operator, a line operator to a line operator, etc. That is, the FDLU preserves the ``geometry'' of the support of an operator: suppose an operator $O(\Omega^d)$ acts on a region $\Omega^d$ of dimension $d\leq D$, then after conjugated by a FDLU, the operator is still supported on a region of dimension $d$.
To prove this statement, note first that due to the ``light-cone structure'' of FDLUs, an operator supported on $\Omega^d$, cannot be mapped to one supported on $\Omega^k$, with $k>d$. The other half of the statement is that a finite-depth circuit cannot map an operator supported on $\Omega^d$ to an operator supported on $\Omega^k$, with $k<d$. We can prove this part by contradiction. Suppose there is a finite-depth circuit $U$ that can map $O(\Omega^d)$ to $O'(\Omega^k)$ with $k<d$, then the inverse $U^{-1}$ maps $O'(\Omega^k)$ to $O(\Omega^d)$. However, this is not possible, since $U^{-1}$ is also a finite-depth circuit. 

The statement can also be generalized and applies to finite-depth transformation of quasi-local unitaries (FDqLU) with at most exponentially decaying tails. Such FDqLU preserves the ``geometry'' of the support of an operator, up to an exponentially decaying tail.

Next we focus on the particular FDLUs (or FDqLUs), those that generate emergent symmetries in topological orders. As discussed above, these unitaries leave the bulk in the ground state, but can permute boundary states. Later, we will give explicit examples of such unitaries.

The ``geometry-preserving'' property of such symmetry-generating finite-depth circuits leads to our claim above. For any local experiment we perform near one boundary, let us record the sequence of operations by $O(\Omega^d)$, where $d$ is at most $D$, and $\Omega^d$ is composed of a finite number of disconnected components near the boundary. The length of $\Omega^d$ along any dimensions perpendicular to the boundary is finite and independent of the system size. Suppose in the ground state with one gapped boundary condition, denoted as $|\Psi\rangle$, the result of the experiment is
\begin{align}
    \langle \Psi | O(\Omega^d) |\Psi\rangle = x\in \mathbb{C}.
\end{align}
Then consider the same ground state with the other gapped boundary condition related by the FDLU (or FDqLU), which is $U|\Psi\rangle$. Due to the ``geometry''-preserving property of $U$, we can always find another operator, with a support of dimension $d$ also near the boundary,
\begin{align}
    O'({\Omega'}^d)=UO(\Omega^d)U^{-1}.
\end{align}
It is easy to confirm that the experiment performed with the sequence of operations $O'({\Omega'}^d)$ produces the same result $x$ on the state $U|\Psi\rangle$.

\subsubsection{Emergent symmetries from invertible defects}
What is a general rule to discover emergent symmetries of topological order as well as the FDLUs (or FDqLUs)? One way is to identify the invertible codimension-$1$ defects in the topological order. Imagine that on the ground state, we apply the unitary $U$ that generates an emergent symmetry yet restricted within a region $\Omega^D$ of dimension $D$, rather than the entire space. Within $\Omega^D$, the state is still in the ground state (i.e., the reduced density matrix for a region within $\Omega^D$ is unchanged). Yet, on the boundary of $\Omega^D$, a symmetry defect is created. Since the inverse of a FD(q)LU is also a FD(q)LU, the defect created by $U(\Omega^D)$ and that created by $U^{-1}(\Omega^D)$ always cancel out. Such defects are thus always invertible. Reversely, given a codimension-$1$ invertible defect in a topological order since sweeping the defect always generates a global emergent $0$-form symmetry. 

\label{subsubsec:examples}
\subsubsection{Two copies of toric codes in $2d$}\label{subsubsubsec:2toriccodes}
\paragraph{Twin states on the boundary} Our first example is where the topological bulk is given by two copies of toric codes. Let us label the anyons as $\{1,e^{\text{I}},m^{\text{I}},f^{\text{I}}\}\times \{1,e^{\text{II}},m^{\text{II}},f^{\text{II}}\}$, with $e^\text{I},e^{\text{II}}$ the gauge charges, and $m^\text{I},m^{\text{II}}$ the gauge fluxes. 
There are in total $6$ types of gapped boundaries for two copies of $\mathbb{Z}_2$ toric codes. Among them two types are of our interest, 
\begin{itemize}
    \item The ``all flux condensed'' boundary. On this boundary, $m^{\text{I}}, m^{\text{II}}$ and their composite are condensed;
    \item The ``twisted'' boundary. On this boundary, $m^{\text{I}}e^{\text{II}}, m^{\text{II}}e^{\text{I}}$ and their composite are condensed.
\end{itemize}
These two boundary states have only relative distinction.

We choose to write two copies of toric code models on a square lattice, with two types of qubits defined on each edge. If the lattice has no boundary, the Hamiltonian is 
\begin{align}
    H=-\sum_{a=\text{I},\text{II}}\left(\sum_vA_v^{a}+\sum_p B_p^a\right),~~A_v^a=\prod_{e\ni v}X_e^a,~~B_p^a=\prod_{e\in p}Z_e^a.
    \label{eq:2toriccodes}
\end{align}
Now we consider the lattice has a smooth boundary on the bottom, and we could index the vertices on the boundary by $i\in \mathbb{Z}$. The Hamiltonian needs to be modified, 
\begin{align}
    H=&H^{\text{bulk}}+H^{\text{bdry}},\nonumber \\
    H^{\text{bulk}}=&-\sum_{a=\text{I},\text{II}}\left(\sum_{v\not\in \text{~bdry}}A_v^{a}+\sum_p B_p^a\right),~~A_v^a=\prod_{e\ni v}X_e^a,~~B_p^a=\prod_{e\in p}Z_e^a,
    \label{eq:2tc}
\end{align}
and different choices of $H^{\text{bdry}}$ lead to different types of boundary states. In particular,
\begin{itemize}
\item The ``all flux condensed'' boundary.
\begin{align}
    H^{\text{bdry}}_{m^{\text{I}}, m^{\text{II}}}=-\sum_{i\in \text{~bdry}}\left(A_i^{\text{I}}+A_i^{\text{II}}\right).
    \label{eq:m_condensed}
\end{align}
Physically, each boundary term is a ``short string'' operator that creates a pair of $m^{\text{I}}$'s or $m^{\text{II}}$'s near the boundary. 
    \item The ``twisted'' boundary.
\begin{align}
    H^{\text{bdry}}_{m^{\text{I}}e^{\text{II}}, m^{\text{II}}e^{\text{I}}}=&-\sum_{i\in \text{~bdry}}\left(A_i^{\text{I}}Z_{e=\langle i,i+1\rangle}^{\text{II}}+A_i^{\text{II}}Z_{e=\langle i-1,i\rangle}^{\text{I}}\right).
    \label{eq:twisted_m_condensed}
\end{align}
And here each boundary term is a ``short string'' operator that creates a pair of $m^{\text{I}}e^{\text{II}}$'s or $m^{\text{II}}e^{\text{I}}$'s near the bottom boundary.
\end{itemize}
One can see that for either choice of the boundary Hamiltonian, all terms in the full Hamiltonian commute. They form a complete set of stabilizers, the ground state is the single common eigenstate that all stabilizers have eigenvalue $1$. Let us call the ground state with the two boundary conditions as $|\Psi\rangle_{m^{\text{I}}, m^{\text{II}}}$ and $|\Psi\rangle_{m^{\text{I}}e^{\text{II}}, m^{\text{II}}e^{\text{I}}}$, respectively.

There is a single depth circuit in $2D$ that maps one state to the other, $U|\Psi\rangle_{m^{\text{I}}, m^{\text{II}}}=|\Psi\rangle_{m^{\text{I}}e^{\text{II}}, m^{\text{II}}e^{\text{I}}}$,
\begin{align}
    U=\prod_{f}(-1)^{g_{W(f)}^Ig_{N(f)}^{II}+g_{S(f)}^Ig_{E(f)}^{II}},~~g^a=\frac{1+Z^a}{2},~~a=\text{I},\text{II}, 
    \label{eq:1dcluster_pump}
\end{align}
where $W(f),N(f),S(f),E(f)$ represent the edges on the west, north, south and east of the face $f$, respectively. In particular, the unitary maps the Hamiltonian terms near the bottom boundary as $A_i^I\leftrightarrow A_i^I Z_{e=\langle i,i+1\rangle}^{II}, A_i^I\leftrightarrow A_i^{II} Z_{e=\langle i-1,i\rangle}^{I}$, up to bulk Hamiltonian terms.

\paragraph{Emergent $\mathbb{Z}_2$ symmetry and ``gauged SPT'' line defect}
The unitary commutes with the bulk Hamiltonian terms up to themselves. For example, $UA_v^{\text{I}}U^{-1}=A_v^\text{I}B_{p(v)}^{\text{II}}$, where $v$ is the left-bottom vertex of the plaquette $p(v)$. Thus, the unitary preserves the ground state subspace of the topological bulk. And it exchanges the anyons $m^{\text{I}}\leftrightarrow m^{\text{I}}e^{\text{II}}, m^{\text{II}}\leftrightarrow m^{\text{II}}e^{\text{I}}$. Thus, it generates an emergent $\mathbb{Z}_2$ symmetry in the topological bulk. When the unitary is restricted to a disk in the bulk, an invertible line defect known as ``gauged SPT'' defect is created on the boundary of the disk. This name comes from the fact that the line defect alternatively comes from taking a one-dimensional $\mathbb{Z}_2\times \mathbb{Z}_2$ SPT within the two-dimensional ambient trivial $\mathbb{Z}_2\times \mathbb{Z}_2$ symmetric state and gauging the $\mathbb{Z}_2\times \mathbb{Z}_2$ global symmetry.

\paragraph{No absolute distinction} Consider the two copies of toric code with a single gapped boundary with two possible choices: the ``all fluxed condensed'' type, and the ``twisted'' type. Can we tell which type the boundary is with only local experiments near the boundary? The answer is no. Since for any local operation near one type of the boundary, through the FDLU in (\ref{eq:1dcluster_pump}), we can find a local operator near the other type of the boundary. Without declaring \emph{a priori} among four anyons with identical topological properties, which are $m^\text{I}, m^{\text{II}},m^\text{I}e^\text{II}, m^{\text{II}}e^\text{I}$, one cannot unambiguously determine which anyons the local operations are detecting the properties of. 

\begin{figure}
    \centering
    \includegraphics[scale=0.33]{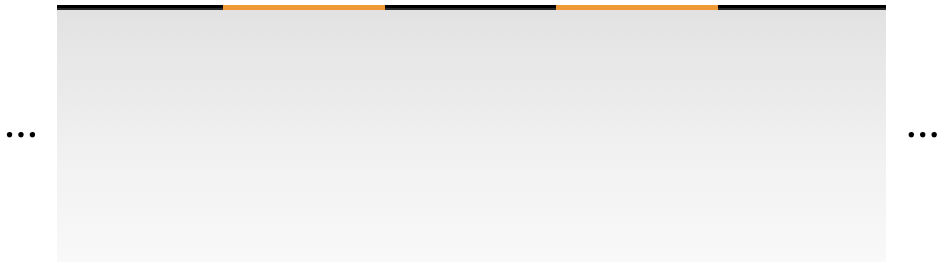}
    \caption{The topological bulk state (say two copies of toric codes) with both types of boundary states in presence. On black segments, one set of anyons (say $m^\text{I}, m^\text{II}$ type) are condensed, and on orange segments, another set of anyons (say $m^\text{I}e^\text{II}, m^\text{II}e^\text{I}$ type) are condensed. Such a system has degenerate ground states which cannot be removed by local perturbations.}
    \label{fig:relative_bdry}
\end{figure}

\paragraph{Relative distinction} Nevertheless, there are ways to confirm that the two types of boundaries are distinct, if we consider the two copies of toric codes with both types of boundary states in presence simultaneously. For example, consider the gapped boundary state shown in Figure~\ref{fig:relative_bdry}. Such a gapped state has a non-local degeneracy. When there are $2N$ segments on the gapped boundary, on which two sets of anyons alternatively condense, there are in total $4^{N-1}$ degenerated states. The degeneracy can never be lifted by local operators either within a segment or a junction between two segments, but only by non-local operators, for example, string operators with endpoints on different segments.

\subsubsection{Toric code in $2d$} 

\paragraph{Twin states on the boundary} Given a toric code ground state in two dimensions, there are two types of gapped boundaries:

\begin{itemize}
    \item The $e$-condensed boundary. On this boundary, the gauge charge, or ``$e$'' anyon, is condensed.
    \item The $m$-condensed boundary. On this boundary, the gauge flux, or ``$m$'' anyon, is condensed. 
\end{itemize}
They serve as our second example of states with only relative distinction. 

\paragraph{$e-m$ exchange symmetry and the line defect}The toric code topological order has an emergent symmetry, the $\mathbb{Z}_2$ symmetry that permutes the $e$-anyon and the $m$-anyon. When we apply the symmetry only to a region $\Omega$, a line defect is generated on $\partial \Omega$. An $e$-anyon passing through the defect becomes an $m$-anyon, and vice versa. Let us consider the toric code model on a square lattice, whose Hamiltonian is as in (\ref{eq:2toriccodes}) but for only one type. Then the FDLU that creates the defect \cite{barkeshli2023codimension} is $U_{\text{pump}}=U_3U_2U_1$, with
\begin{align}    U_1=\prod_{\text{verticle~}e\in \Omega}e^{-\ii\frac{\pi}{4}U_e},~~U_2=\prod_{v\in\Omega}e^{-\ii\frac{\pi}{4}Y_{e_{1}(v)}Y_{e_{2}(v)}},~~U_3=\prod_{\text{horizontal~}e\in \Omega}e^{-\ii\frac{\pi}{4}U_e}.
   \label{eq:em_exchange_pump}
\end{align}
Here, $U_e$ is the short string creating $f=e\times m$ anyons,
\begin{align}
U_e=\vcenter{\hbox{\includegraphics[scale=1.0]{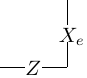}}}~~\text{or}~~\vcenter{\hbox{\includegraphics[scale=1.0]{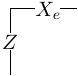}}}.
\label{eq:Ue}
\end{align}
An example of the unitary is illustrated in Figure~\ref{fig:boundary_majorana_chain_pump}. The unitary $U_\text{pump}$ exchange bulk Hamiltonian terms $A_v\leftrightarrow B_{p(v)}$, where $v$ is the left-bottom vertex of $p(v)$, and thus exchange $e$ and $m$ anyons.
The intuition for deriving this unitary is that one can start with the unitary that pumps a Kitaev chain on a loop in a trivial fermion insulator, which is composed of fermion bilinears. Then through a $2d$ \emph{bosonization}, (in other words, to gauge the fermion parity symmetry,) the trivial fermionic state is mapped to the toric code state, and the unitary becomes the one given above.

On the line defect, the bound state of gauge charge and gauge flux $f=e\times m$, which is a self-fermion, is condensed. To see this, one can apply the unitary $U(\Omega)$ on the set of stabilizers that define the toric code ground state by conjugation. In the new set of stabilizers, only the star terms $A_v=\prod_{e\in v}X_e$ and the plaquette terms $B_f=\prod_{e\in f}Z_e$ for $v$ and $f$ near $\partial \Omega$ are replaced by short open string operators that create pairs of $f$.

\begin{figure}[ht]
    \centering
      \includegraphics[scale=0.43]{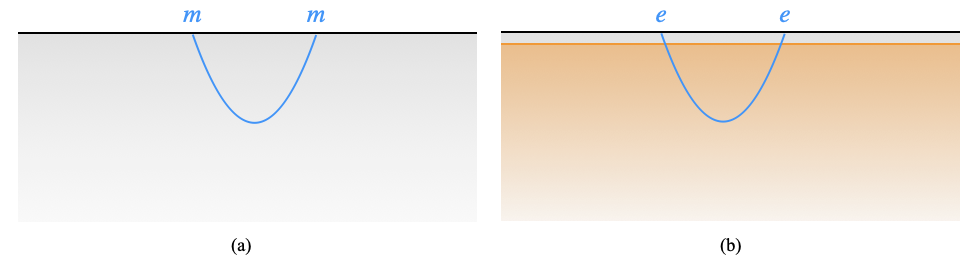}
    \caption{(a) Toric code with a $m$-condensed boundary. (b) Toric code with a $e$-condensed boundary, which is obtained from (a) by applying the FDLU in (\ref{eq:em_exchange_pump}).}
    \label{fig:toric_code_bdry}
\end{figure}

\paragraph{No absolute distinction} The toric code ground state with two types of boundaries are related by a two-dimensional FDLU in (\ref{eq:em_exchange_pump}), as illustrated in Figure~\ref{fig:toric_code_bdry}. As a result, for any local operation near one type of the boundary, say the $m$-condensed boundary in (a), there is a corresponding operation near the other type of the boundary, say the $e$-condensed boundary in (b), that is obtained from the previous one by conjugating the FDLU in (\ref{eq:em_exchange_pump}). In other words, if we can only perform local experiments near a single type of boundary, we cannot identify which type the boundary is, without the reference of bulk excitations. For example, a reference could mean making the choice that the presence of $e$-anyon is detected by the violation of what local term, either a star term or a plaquette term.

\begin{figure}[ht]
    \centering
    \includegraphics[scale=0.33]{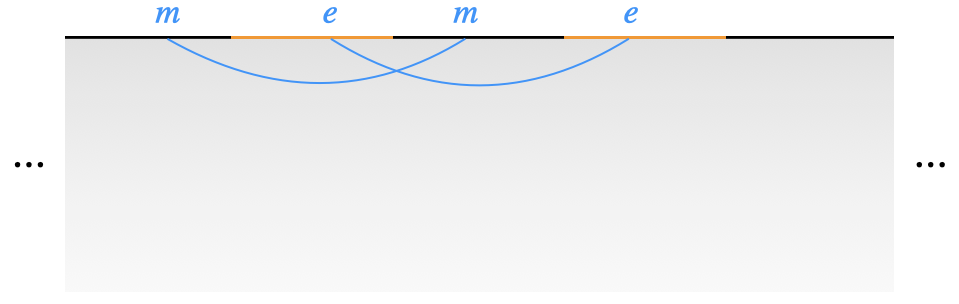}
    \caption{The toric code with both types of boundary states in presence. On black segments, one type of anyons (say $m$ type) are condensed, and on orange segments, another type of anyons (say $e$ type) are condensed. Such toric code model has degenerate ground states. And the degeneracy cannot be lifted by any local perturbation near the domain walls between segments.}
    \label{fig:toric_code_relative_bdry}
\end{figure}

\paragraph{Relative distinction} Even though within a single gapped boundary, we cannot detect the type of gapped boundary through local experiments, yet there are indeed two distinct types of boundaries, as we can detect the difference between them, when the two are both present. For example, consider the gapped boundary state shown in Figure~\ref{fig:toric_code_relative_bdry}. Such a gapped state has a non-local degeneracy. When there are $2N$ segments on the gapped boundary, on which $e$ and $m$ alternatively condense, there are in total $2^{N-1}$ degenerated states. The degeneracy can never be lifted by local operators either within a segment or a junction betewen two segments, but only by non-local operators, for example, string operators with endpoints on different segments.

\subsubsection{Toric code in $3d$}
\paragraph{Twin states on the boundary} Consider the toric code ground state in three dimensions, there are also two types of gapped boundaries that only have relative distinction: 
\begin{itemize}
    \item The ``smooth'' boundary. On this boundary, the gauge flux loop, denoted as $m_{\text{loop}}$ is condensed. This boundary is the analog of the $m$ (gauge flux) condensed boundary of toric code model in two dimensions. Furthermore, 
    the gauge flux loop is allowed to break up into a string, if the endpoints terminate at the boundary. Nevertheless, we emphasize that the end points of the gauge flux loop that terminate on the boundary always costs energy proportional to the length of the string. Thus, there is a string tension between two terminations attached with a single $m_{\text{loop}}$.
    \item The twisted ``deconfined'' boundary, or the twisted ``smooth'' boundary. This boundary shares many similarities with the ``smooth'' boundary. The gauge flux loop is condensed, and can break up into a string which terminates on the boundary. The termination point of the gauge string is not deconfined point-like excitation, due to the string tension between two terminations attached with a single $m$-string. The key difference is that on the twisted boundary, the end point of the gauge flux string has ``semionic statistics'', which could be shown with the effective topological action. \cite{zhao2022string,ji2023boundary}
    Certainly, since the end point is not deconfined, one needs to declare the meaning of braiding statistics of the end points.
\end{itemize}

\paragraph{Emergent $\mathbb{Z}_2$ symmetry and ``gauged $\mathbb{Z}_2$ SPT'' defect } In three-dimensional toric code, there is an invertible codimension-1 defect, the so-called ``gauged $\mathbb{Z}_2$ SPT defect''. More precisely, the defect is obtained by first embedding a codimension-1 defect containing the Levin-Gu SPT state into a three-dimensional $\mathbb{Z}_2$ paramagnet. Then, after gauging the $\mathbb{Z}_2$ symmetry, the codimension-1 defect becomes the surface defect in the toric code. Explicitly, the following unitary operator acting on a volume $\mathcal{V}$ creates the defect on the surface $\partial \mathcal{V}$,
\begin{align}
    U(\mathcal{V})=\prod_{\tetra_{0123}\in \mathcal{V}}e^{\ii \frac{\pi}{8}\left(1+\sum_{i\neq j}(-1)^{i+j}Z_{ij}+Z_{01}Z_{23}\right)},
    \label{eq:TC3d_unitary}
\end{align}
where $0,1,2,3$ labels the four vertices of a tetrahedron, or rather a branched $3$-simplex.\cite{ji2023boundary}

As before, the invertible codimension-1 defect leads to an emergent $0$-form symmetry in the topological phase. The unitary when acting on the whole space without a boundary generates a $\mathbb{Z}_2$ symmetry in the ground state subspace. To see this, the Hamiltonian local terms of the toric code model gives rise to a set of stabilizers. The unitary commutes with the set of stabilizers, up to a basis transformation of the generating stabilizers. Thus, the ground state subspace is invariant when acted by the unitary. Nevertheless, the unitary acts non-trivially within the subspace.

\paragraph{No absolute distinction} The $\mathbb{Z}_2$ symmetry when acting on a system with boundary maps between the two types of smooth boundaries. Because of this, there cannot be \emph{local experiment} merely near the boundary that can distinguish the two boundary conditions. For example, a natural guess of such an experiment is to detect the ``self-statistics'' of the endpoint of the flux loop. In order to extract a meaningful statistical phase, a braiding experiment should be performed with a large separation between the different excitations. Since the excitations are confined, such an experiment cannot be performed, so this statistical phase cannot be extracted.

\begin{figure}[ht]
    \centering
    \includegraphics[scale=0.4]{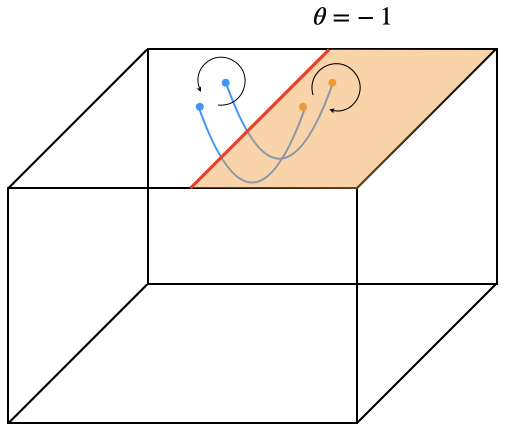}
    \caption{The toric code model in 3D, with the ``smooth'' boundary state and the ``twisted smooth'' boundary state meet at a 1d domain wall.}
    \label{fig:3d_tc_relative_difference}
\end{figure}

\paragraph{Relative distinction} 
However, the ``smooth'' boundary and the twisted ``smooth'' boundary are two different types of gapped boundaries. We can tell the difference when both are present and separated by a domain wall, as illustrated in Figure~\ref{fig:3d_tc_relative_difference}. First we consider pushing two fluxes partially to the boundary and let the two end points of each flux string to be on different sides of the domain wall. Next we shrink one flux string slightly, let it pass under the other, then enlarge it and let it pass over the other and back to its original location. In the end, equivalently the endpoint of one string braids with one endpoint of the other string by $2\pi$, and similarly for the other pair of endpoints. The statistical phase accumulation through this process is $-1$, and cannot be removed by applying any local unitaries along the strings and the 1d domain wall. In contrast, if considering the same process of two flux strings without the 1d domain wall, there is no non-trivial statistical phase. Thus, the non-trivial statistical phase signals the relative distinction between the two boundary states and the presence of a non-trivial domain wall between the two. This is similar to the 3-loop braiding process in the bulk of a $3+1$D topological order\cite{Wang2014,Jiang2014}, only truncated by the boundary. 

\subsubsection{Fermionic toric code in $3d$}\label{subsubsec:fTC}
In three dimensions, another non-trivial topological model is the fermionic version of the toric code, where the topological point excitation, or the $\mathbb{Z}_2$ gauge charge, is fermionic instead of bosonic. The Hamiltonian for the fermionic Toric code on a cubic lattic with one qubit per edge is 
\begin{align}
    H = -\sum_v A_v -\sum_f G_f,~~~A_v=\prod_{e\ni v} X_e,
    \label{eq:fTC}
\end{align}
where 
\begin{align}
    G_f=\vcenter{\hbox{\includegraphics[scale=1.0]{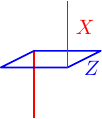}}}~\text{or}~~\vcenter{\hbox{\includegraphics[scale=1.0]{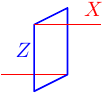}}} ~\text{or}~~ \vcenter{\hbox{\includegraphics[scale=1.0]{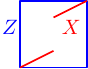}}}~~,
\end{align}
depending on the orientation of $f\in \{xy,yz,zx\}$.

\paragraph{Twin states on the boundary} Given the fTC ground state in three dimensions, there are also two types of gapped boundary states that are only relatively distinguishable:
\begin{itemize}
    \item The ``smooth'' boundary. The gauge flux loop is condensed on this boundary. If the endpoints of the loop terminates on the boundary, the endpoint has bosonic statistics. 
    \item The twisted ``smooth'' boundary. The gauge flux loop is also condensed on this boundary. Nevertheless, the endpoint of flux loop carries a non-abelian Majorana mode, which we can label by $\sigma$. When we reconnect or fuse the two endpoints of a loop, there is a fusion channel that give rise to a complex fermion $\psi$, based on the fusion rule $\sigma\times \sigma = 1+\psi$.
\end{itemize}
\paragraph{``Gauged $p+ip$'' defect and emergent symmetry} The fermionic toric code (fTC) has an emergent $0$-form symmetry. To see this, recall that in fTC, a ``gauged $p+ip$ superconducting phase'' can be pumped on a closed surface by a unitary operator acting on the volume enclosed by the surface \cite{fidkowski2023pumping}. In this way, an invertible defect, sometimes dubbed ``gauged $p+ip$ defect'' is created. When we sweep such a defect across the whole three-dimensional space, the ground state subspace is intact. More precisely, for a fTC on a $3d$ lattice without a boundary, the unitary operator acting on the whole space generates a 0-form symmetry within the ground space subspace. For example, the unitary acts as a logical $CCZ$ gate on the three qubits given by the fTC ground states on a $T^3$ torus. The unitary is only quasi-local, in the sense that if we take the unitary as a finite time evolution of a Hermitian operator, terms in the Hermitian operator are allowed to have exponentially decaying tails. The unitary can thus be taken as a finite-depth transformation of quasi-local unitaries (FDqLU). On the excited state, the unitary acts non-trivially. In particular, under the unitary, a flux loop will be dressed by a Majorana chain.

The unitary when acting on a $3D$ lattice with a boundary, keeps the bulk in the fTC ground state, but will transform the ``smooth'' boundary state into the twisted ``smooth'' boundary. Especially, under the action of the unitary, since the flux string is decorated by a Majorana chain, and the endpoint of the string when landing on the boundary carries a non-Abelian Majorana mode. 

\paragraph{No absolute distinction} The smooth boundary and the twisted smooth boundary, as they are related by a FDqLU, also have no absolute distinction. Any local operation near one boundary corresponds to a quasi-local operation near the other boundary, obtained by conjugating the FDqLU. Local experiments near a single type of boundary without the reference of bulk excitations are not sufficient to determine the boundary type.

\begin{figure}[H]
    \centering
    \includegraphics[scale=0.4]{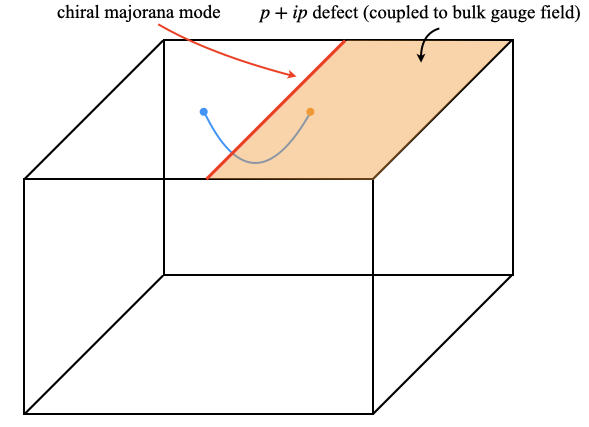}
    \caption{The fermionic toric code model in 3D, with the ``smooth'' boundary state and the twisted ``smooth'' boundary state meet at a 1d domain wall.}
    \label{fig:3d_ftc_relative_difference}
\end{figure}

\paragraph{Relative distinction} To distinguish the two types of boundaries, we can consider the state in Figure~\ref{fig:3d_ftc_relative_difference}, where the two boundary states share a domain wall. There is a chiral Majorana mode running along the $1d$ domain wall. To see this, consider the scenario that a gauge flux loop is partially pushed to the boundary, and becomes an open flux string. And consider the case that one endpoint is on the ``smooth'' boundary, while the other is on the twisted ``smooth'' boundary. Then only one endpoint carries a Majorana mode. Yet during the process, the dimension of the total Hilbert space remains to be an integer. That means there is a Majorana mode on the boundary. Such a non-trivial $1d$ domain wall signals that the two boundary states neighboring the domain wall are distinct.  

\subsection{Invertible phases in Topological Holography}
\label{sec:Invert_symTFT}

In this subsection, we obtain invertible phases using the ``sandwich construction'' shown in Figure~\ref{fig:sandwich}. We will obtain not only the ground state, but also LEEs in the phase. Because of the relative distinction between certain pairs of boundary states given a topological bulk, we show an equivalence between the low entanglement excitations in a non-trivial invertible phase and that of a trivial one. 

\begin{figure}[th]
    \centering
   \includegraphics[width=0.98\linewidth]{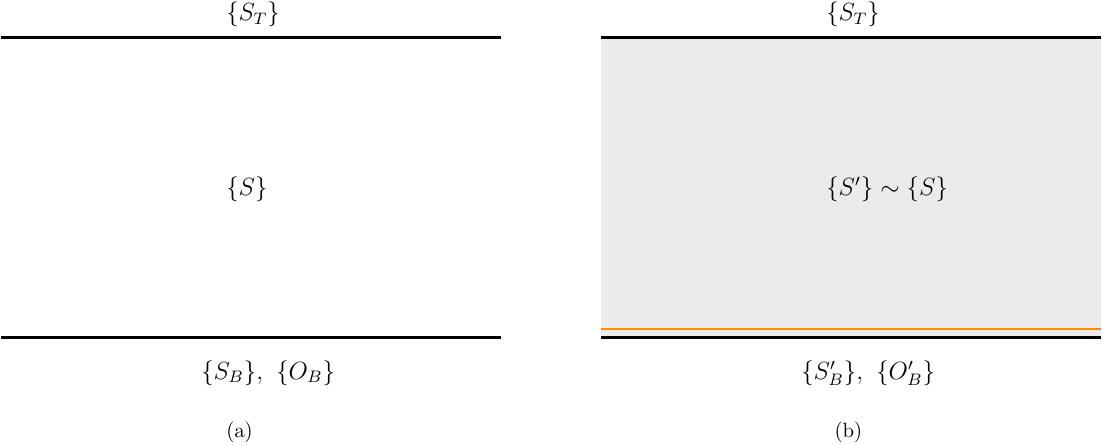}
    \caption{Schematic illustration for the ``sandwich constructions'' of phases with relative difference. (a) A ``sandwich'' for a gapped phase:  the choice of topological ordered bulk and topological boundaries can be specified by a commuting projector Hamiltonian with bulk terms $\{S\}$, and boundary terms $\{S_{T}\}$, $\{S_{B}\}$. Any operators $O_B$ near the bottom boundary that commute with all bulk and top boundary terms $\{S,S_T\}$, but not all bottom boundary terms $S_B$ create excitations near the bottom boundary. (b) The ``sandwich'' for a gapped phase that differs from (a) by a $D$-dimensional invertible defect (orange line). It can be obtained from the sandwich (a) by conjugating all terms $\{S_T, S, S_B, O_B\}$ by the unitary if the top boundary $\{S_T\}$ is invariant under the conjugation. The excitation created by $O_B$ in (a) has a correspondence in (b) created by $O_B'=UO_B'U^{-1}$.}
    \label{fig:schematic_sandwiches}
\end{figure}

A ``sandwich'' is a topological order in $D+1$ spatial dimensions with two parallel boundaries: the top (topological) boundary is chosen to be a gapped boundary, and the bottom (dynamical) boundary can be in any boundary condition. We illustrate the construction schematically in Figure~\ref{fig:schematic_sandwiches}(a) using a commuting projector Hamiltonian. The purpose is that after we ``squash'' the sandwich, we obtain a quasi-$D$-dimensional system with a global symmetry. The symmetry depends on the topological boundary condition we begin with. 

Let us sketch the generic steps we use to obtain invertible phases based on the construction. Given a topological order in $D+1$ spatial dimensions, we construct two ``sandwiches'' with identical top topological boundaries but distinct gapped dynamical boundaries, differentiated by an invertible defect. Thus, after ``squashing'', the two $D$-dimensional systems have the same global symmetry. One realizes the trivial invertible phase; the other, as we will show, becomes a non-trivial invertible phase. And more importantly, we can establish the bijection between symmetric operators, such as those creating low entanglement excitations in the trivial phase and those in the non-trivial invertible phase. 

We illustrate the commuting projector Hamiltonian terms $\{S_T\}, \{S\}, \{S_B\}$ at the top boundary, in the bulk and at the bottom boundary, respectively and excitation creation operators $\{O_B\}$ for one ``sandwich'' in Figure~\ref{fig:schematic_sandwiches}(a). These terms after conjugation by the pump unitary $U$ creating an invertible defect become the terms in the other sandwich in Figure~\ref{fig:schematic_sandwiches}(b). Since the pump unitary keeps the bulk ground state wavefunction invariant, the bulk Hamiltonian terms $\{S'=USU^{-1}\}$ can be replaced by $\{S\}$. In most cases, the top boundary terms $\{S_T\}$ are also invariant under the conjugation by $U$. The bottom boundary terms $\{S_B'=US_BU^{-1}\}$ and $\{O_B'=UO_BU^{-1}\}$ have the property that they still commute with all bulk terms $\{S\}$, and thus define the ``twisted'' bottom boundary condition and excitations above. 

We remark that in some cases, the top boundary terms $\{S_T\}$ are not invariant under the conjugation of $U$. This includes the Kitaev chain example and the $p+ip$ example. For these two cases, we need to establish the equivalence in a slightly different way. For all other cases, $U$ acts effectively as a locality preserving unitary in the quasi $D$-dimensional system that maps a trivial invertible phase to a non-trivial one. It leads to a bijection between symmetric operators in the two phases and correspondingly a bijection between the LEEs.

\label{sec:example}
\subsubsection{$\mathbb{Z}_2\times\mathbb{Z}_2$ SPT in $1D$}
\begin{figure}[ht]
    \centering
    \includegraphics[scale=0.4]{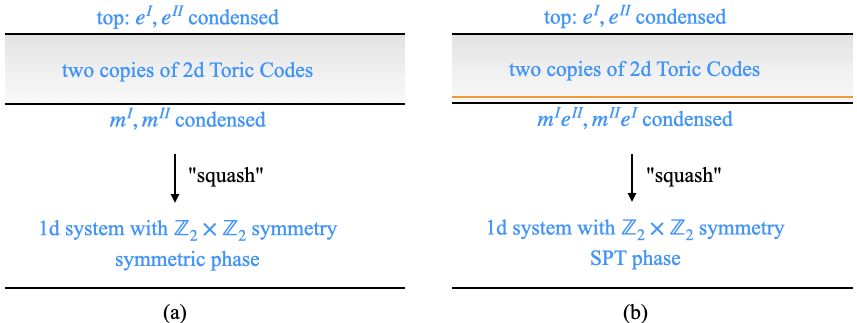}
    \caption{``Sandwich construction'' from two copies of toric codes in 2D. The top (topological) boundary is chosen to be the gauge charge condensed boundary. The bottom (dynamical) boundary is chosen to be (a) flux condensed boundary (b) the ``twisted'' flux condensed boundary.}
    \label{fig:sandwich_SPT1D}
\end{figure}

Let us explain the ``sandwich'' on the lattice. Usually, it is convenient to describe the bulk and the top topological boundary in terms of a fixed-point Hamiltonian. As a simple example, we consider one-dimensional system with $\mathbb{Z}_2\times \mathbb{Z}_2$ symmetry. The corresponding ``sandwich'' considers two copies of toric codes on a strip as in Figure~\ref{fig:sandwich_SPT1D}. The bulk part of the fixed point Hamiltonian is the same as in Eq.~(\ref{eq:2tc}) while the top boundary part can be chosen to be $-\sum_{\text{top edge }e}(Z_e^I+Z_e^{II})$, which means that $e$-anyons are condensed on the top boundary.
For the dynamical bottom boundary, in general, we are free to choose any Hamiltonian, in which each term is local and near the bottom boundary and commutes with the terms in the fixed-point Hamiltonian. 

\begin{figure}[ht]
    \centering
    \includegraphics[scale=0.7]{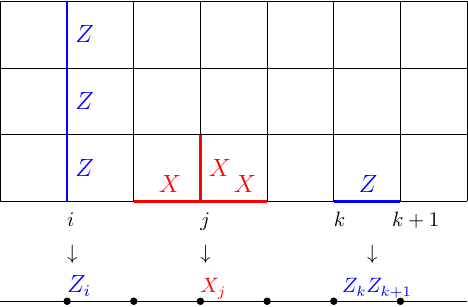}
    \caption{Operator mapping from the ``sandwich'' to a quasi-$1D$ system. In the ``sandwich'', the bulk is the toric code ground state while $e$ anyons are condensed on the top boundary.}
    \label{fig:operator_reduction}
\end{figure}

The first claim is that such a model on a strip with two boundaries, in the limit that the strip is thin, is equivalent to a quasi-1D system with a global symmetry. Indeed, a mapping between operator algebra in the 2D system on a thin strip and that in a 1D system can be obtained as follows. Within the ground state subspace stabilized by the Hamiltonian terms in the bulk and on the top boundary, the logical Pauli-$Z$ and Pauli-$X$ operators (of either type $I$ or type $II$, whose index we suppress) are given by a vertical $Z$-type string operator and a $X$-type of truncated star operator, as shown in Figure~\ref{fig:operator_reduction}.

In particular, the global symmetry generator $\prod_j X_j$ (of either type $I$ or type $II$ with index suppressed) in the 1D system is mapped from a Wilson loop operator along the long direction in the strip,
\begin{align}
    W_m(C^\vee)=\prod_{e\in C^\vee}X_e \rightarrow \prod_j X_j, \label{eq:Wilsonloop}
\end{align}
where $C^\vee$ is a loop on the dual lattice along the long direction of the strip.

Now we focus on the dynamical bottom boundary and make two choices. In one choice, the boundary has the Hamiltonian given in Eq.~(\ref{eq:m_condensed}), also illustrated in Figure~\ref{fig:sandwich_SPT1D}(a). In the other choice, the boundary Hamiltonian is given in Eq.~(\ref{eq:twisted_m_condensed}), and illustrated in Figure~\ref{fig:sandwich_SPT1D}(b). As we discussed in Section \ref{subsubsec:examples}, the two choices of bottom boundaries differ by an gauged-SPT defect, which can be created by a 2d FDLU.

With either choice, the sandwich can be reduced to a quasi-1D system with $\mathbb{Z}_2\times\mathbb{Z}_2$ symmetry, since the symmetry generators commute with the boundary Hamiltonian, as well as the fixed-point Hamiltonian in the bulk and on the top boundary. 

Indeed, applying the operator map, it is obvious to see that the two types of bottom boundary Hamilonians (\ref{eq:m_condensed}) and (\ref{eq:twisted_m_condensed}) - the only dynamical part of the sandwich model - are mapped to 
\begin{align}
    H^{\text{bdry}}_{m^I,m^{II}}\rightarrow& -\sum_i (X_i^I+X_i^{II} )\nonumber \\
    H^{\text{bdry}}_{m^Ie^{II},m^{II}e^{I}}\rightarrow& -\sum_i (X_i^I Z_{i}^{II}Z_{i+1}^{II}+Z_{i-1}^{I}Z_i^I X_i^{II} ),
\end{align}
where $UH^{\text{bdry}}_{m^I,m^{II}}U^{-1}\simeq H^{\text{bdry}}_{m^Ie^{II},m^{II}e^{I}}$ with the $2D$ FDLU $U$ given in Eq.~(\ref{eq:1dcluster_pump}), up to bulk Hamiltonian terms, as we show in Section \ref{subsubsec:examples}.
Thus, through ``squashing two sandwiches'', we recover the fixed-point Hamiltonian of the paramagnetic state and that of the SPT state in $1D$.

The second claim is that there is a bijection between the symmetric operators in the paramagnetic phase and those in the SPT phase. Recall that the two boundary Hamiltonians are related by the conjugation by the 2d FDLU $U$ or its inverse $U^{-1}$. In fact, any operator near the bottom boundary and commuting with the bulk Hamiltonian terms, as well as the Wilson loop operator in Eq.~(\ref{eq:Wilsonloop}) in the ``sandwich'' in Figure~\ref{fig:sandwich_SPT1D}(a), can be conjugated by $U$ and becomes an operator still near the bottom boundary in the ``sandwich'' in Figure~\ref{fig:sandwich_SPT1D}(b); and similarly there is a mapping in reverse direction. Then through the operator mapping, the operator and its conjugated one become operators acting on the trivial and non-trivial symmetric phase, respectively. Thus, a bijection is established. 

Let us also remark to what extent the FDLU $U$ can be taken as a $1d$ locality preserving unitary when the sandwich or the strip geometry is taken as a quasi-1d system. First, $U$ is a quasi-$1d$ FDLU. Second, there is a $\mathbb{Z}_2\times \mathbb{Z}_2$ symmetry generated by Wilson loop operators $W_{m^{\text{I}}}(C^\vee)$ and $W_{m^{\text{II}}}(C^\vee)$ that take form as in Eq.~(\ref{eq:Wilsonloop}), within the subspace that the bulk of the sandwich is in the ground state of two copies of toric codes. The quasi-1d FDLU commutes with the symmetry generators \emph{only} in the subspace that $\prod_{e\in p}X_e^{\text{I}}=1$, $\prod_{e\in p}X_e^{\text{II}}=1$ for all plaquette $p$. Within the subspace, $U$ is a non-trivial symmetric locality preserving unitary.

\subsubsection{Kitaev chain}
\begin{figure}[ht]
    \centering
    \includegraphics[scale=0.4]{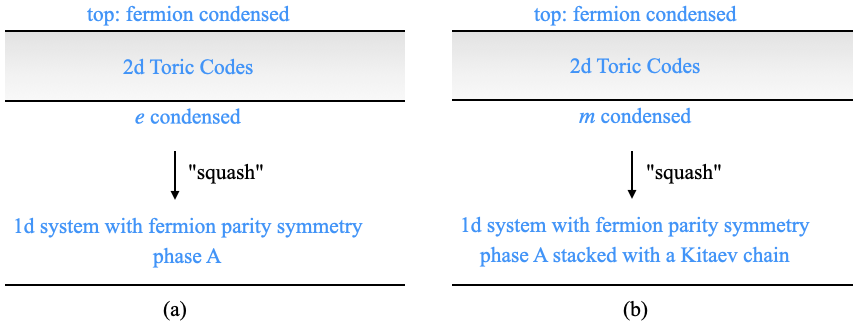}
    \caption{``Sandwich construction'' from the toric code in 2D. The top (topological) boundary is chosen to be the fermion condensed boundary (see Appendix \ref{app:fcondensedbdry}). The bottom (dynamical) boundary is chosen to be (a) gauge charge condensed boundary (b) gauge flux condensed boundary.}
    \label{fig:sandwich_Kitaev}
\end{figure}

In this example, we consider the bijection of LEEs preserving fermion parity symmetry in the trivial and non-trivial fermionic invertible phase. The ``sandwich construction'' for fermionic models with total fermion parity symmetry in one dimension is to start with the toric code in two dimensions as the bulk, and a top boundary where fermions are condensed after paired with physical fermions. A discussion about the fermion condensed boundary with explicit Hamiltonians is in Appendix \ref{app:fcondensedbdry}. Let us consider such ``sandwich constructions'' with two choices of bottom boundaries. Schematically, they are shown in Figure~\ref{fig:sandwich_Kitaev}. The explicit Hamiltonians are shown in Figure~\ref{fig:sandwich_f_tc_em}.
\paragraph{Two ``sandwiches''} The generic approach to map between the two sandwiches using pump unitaries as outlined at the beginning of Section 
\ref{sec:Invert_symTFT} does not apply immediately in this Kitaev chain example. To understand this, suppose we apply the ``gauged Kitaev chain'' pump unitary to the sandwich (a) in Figure~\ref{fig:sandwich_Kitaev}, the top fermion-condensed boundary is not invariant, but changed to a twisted version of the fermion condensed boundary, which we discuss in Appendix \ref{app:fcondensedbdry}. The fermion condensed boundary and its twisted version are shown in Figure~\ref{fig:ftop_stabilizers}, (see also Figure~\ref{fig:ftopboundary2} for their fermionized version). This action of pump unitary is found explicitly in Eq.~(\ref{eq:twisted_fboundary}) and (\ref{eq:boundary_majorana_chain_pump}). Since the top fermion-condensed boundary does not absorb a ``gauged Kitaev chain'' line-defect, the pump operator creates two line-defects, while the sandwiches in Figure~\ref{fig:sandwich_Kitaev} with Hamiltonians given in Figure~\ref{fig:sandwich_f_tc_em} differ by a single line defect. 

Nevertheless, we can instead using a sequential circuit only near the bottom boundary to create a single line-defect, and thus maps between the two sandwiches. We give the circuit in Appendix \ref{subsubsec:Kitaev_chain_sandwich}. Within the subspace that Wilson loop operators $W_e (C)=1, W_m(C)=1$ for loops $C$ along the direction parallel to the boundaries, the circuit is locality preserving.

\begin{figure}[ht]
    \centering
    \includegraphics[scale=0.85]{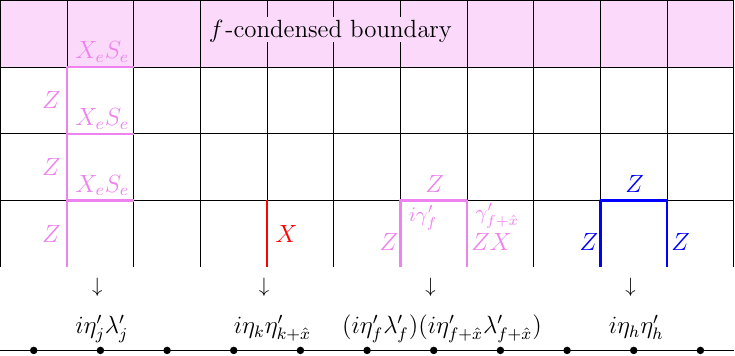}
    \caption{Operator mapping from the ``sandwich'' to a quasi-$1D$ fermionic system.}
    \label{fig:operator_reduction_fermion}
\end{figure}

\paragraph{Dimension reduction, fermion parity symmetry and boundary operator algebra} Sandwiches with $2D$ toric code ground state as the bulk state and the fermion-condensed top boundary are equivalent to quasi-$1D$ systems with a global fermion parity symmetry. We can identify the operator mapping from the ``sandwich'' to the fermionic quasi-$1D$ system with two flavors of fermions per site, as we show in Figure~\ref{fig:operator_reduction_fermion}. Within the ground state subspace stabilized by the Hamiltonian terms in the bulk and on the top boundary, bosonic operators near the bottom boundary are mapped to operators of two flavors of Majorana fermions, which we call $\eta$ and $\eta'$ that compose into a complex fermion; similarly, operators of physical fermions near the bottom boundary are mapped to the other two flavors of Majorana fermions, which we call $\lambda$ and $\lambda'$.

In particular, from the operator mapping, we can derive that the total fermion parity symmetry operator of one flavor, $P^\eta=\prod_{j}(\ii \eta_j\eta_j')$ is mapped from a Wilson loop operator parallel to the boundary, of either $e$-type or $m$-type. More explicitly,
\begin{align}
    W_e(C)=\prod_{e\in C}Z_e\rightarrow P^\eta,~~W_m(C)=\prod_{e\in C^\vee} X_e\rightarrow P^\eta,
\end{align}
up to a numerical phase of either $1$ or $-1$, where $C$ and $C^\vee$ is a loop that winds the ``sandwich'' along the horizontal direction on the lattice and on the dual lattice, respectively.

Applying this operator mapping, one can find that the bottom boundary Hamiltonians shown in Figure~\ref{fig:sandwich_f_tc_em} are mapped to fermionic Hamiltonians in 1D,
\begin{align}
    H^{\text{bdry}}_{e}\rightarrow -\sum_{j}\ii \eta_j\eta_j',\nonumber\\
    H^{\text{bdry}}_{m}\rightarrow -\sum_{j}\ii \eta_j\eta_{j+1}',
\end{align}
where one describes a trivial superconducting phase, while the other describes the topological superconducting phase \cite{kitaev2001unpaired}. 

\paragraph{Bijection of symmetric LEEs of two sandwiches} The sequential circuit in Eq.~(\ref{eq:boundary_circuit}) becomes a sequence of majorana swapping operators under the operator mapping in Figure~\ref{fig:operator_reduction_fermion},
\begin{align}
    R(i\eta_N\eta_N') R(i\eta_N\eta_1')\cdots R(i\eta_j\eta_j')R(i\eta_j\eta_{j+1}')\cdots R(i\eta_1\eta_1')R(i\eta_1 \eta_2'),
    \label{eq:majorana_swaps}
\end{align}
where $R(O)=e^{\ii \frac{\pi}{4}O}$.
which effectively translates all majoranas $\eta_j'\rightarrow \eta_{j}, \eta_j\rightarrow \eta_{j+1}'$, within the fermion parity even sector $P_\eta=1$. Under this unitary of majorana swapping operators, any LEEs created by local operators preserving fermion parity in the trivial phase are thus mapped to LEEs in the topological superconducting phase, and vice versa.

\subsubsection{$\mathbb{Z}_2$ SPT in $2D$}
\begin{figure}[ht]
    \centering
    \includegraphics[scale=0.4]{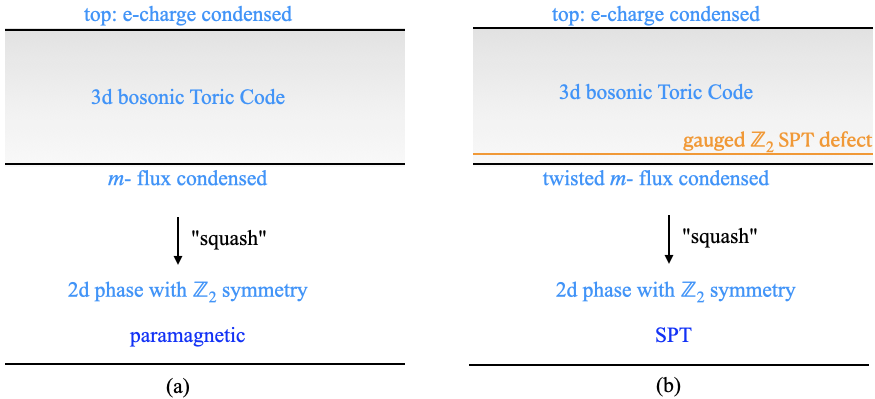}
    \caption{``Sandwich construction'' from the toric code in 3D. The top (topological) boundary is chosen to be the gauge charge condensed boundary. The bottom (dynamical) boundary is chosen to be (a) flux condensed boundary (b) fluxed condensed boundary stacked with the ``gauged $\mathbb{Z}_2$ SPT'' defect, or the ``twisted'' flux condensed boundary. }
    \label{fig:sandwich_SPT2D}
\end{figure}

To illustrate a higher-dimensional example, we examine the bijection of operators between the trivial and nontrivial $\mathbb{Z}_2$ SPT states in two dimensions. Using the ``sandwich construction'', we start with a three-dimensional toric code as the bulk and a top boundary where gauge charges condense. Figure~\ref{fig:sandwich_SPT2D} depicts the two states, which differ by the choice of bottom boundary: either the $m$-flux condensed boundary or the twisted boundary, as described in Section \ref{subsubsec:examples}. After dimension reduction, the two ``sandwiches'' correspond to distinct $\mathbb{Z}_2$ symmetric states. The two-dimensional symmetry generator $\prod_j X_j$ is mapped from a topological surface operator parallel to the boundaries, representing the world history of a flux loop:
\begin{align}
    \prod_{e\perp \mathcal{A}^\vee}X_e\rightarrow\prod_j X_j, \label{eq:surface_op_flux_loop}
\end{align}
where $\mathcal{A}^\vee$ is a surface on the dual lattice parallel to the boundaries.

Since the two boundaries are related by a $3d$ FDLU shown in (\ref{eq:TC3d_unitary}) and differ only relatively, as in previous examples, there exists a bijection between operators near the bottom boundaries that commute with both the topological bulk and the topological surface operator in (\ref{eq:surface_op_flux_loop}). That ensures a bijection between symmetric operator algebras in the two distinct symmetric phases after dimension reduction.

\subsubsection{$p+ip$ superconductor}
Let us also establish the bijection between LEEs in the two dimensional trivial superconducting state and the $p$-wave topological superconducting state. 

\begin{figure}[ht]
    \centering
    \includegraphics[scale=0.4]{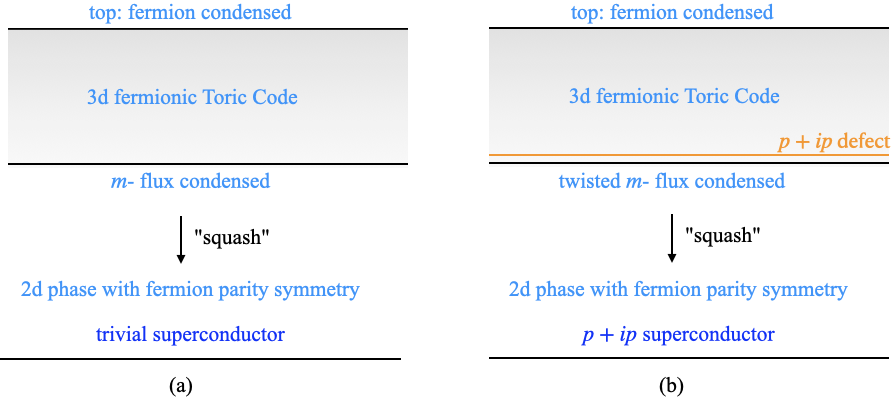}
    \caption{``Sandwich construction'' from the fermionic toric code. The top (topological) boundary is chosen to be the fermion condensed boundary (analogous to the $1d$ boundary shown in Appendix \ref{app:fcondensedbdry}). The bottom (dynamical) boundary is chosen to be (a) flux condensed boundary (b) fluxed condensed boundary stacked with the invertible $p+ip$ defect, or the ``twisted'' flux condensed boundary.}
    \label{fig:sandwich_p+ip}
\end{figure}

At the level of topological quantum field theories, there is still a ``sandwich construction'' that gives rise to the spin TQFT whose ground state is the $p+ip$ superconducting state, as we illustrate in Fig \ref{fig:sandwich_p+ip} (b). To produce the ``sandwich construction'' at the lattice level, a non-trivial part is to obtain the twisted $m$-flux condensed boundary. We can construct it this way. We begin with a $m$-flux condensed boundary. Then we attach a layer of $p+ip$ supercondutor onto the boundary, and let the fermions couple with the gauge field of the bulk fermionic toric code model. 

\begin{figure}[ht]
    \centering
    \includegraphics[width=\linewidth]{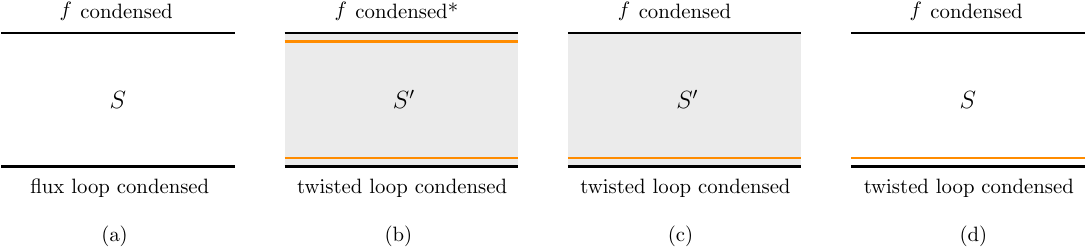}
    \caption{Sandwich constructions that reduce to $2D$ trivial superconductor [(a) and (b)] and $2D$ $p+ip$ superconductor [(c) and (d)]. (a) A simple sandwich equivalent to a $2D$ fermionic state with total fermion parity symmetry; (b) An auxiliary construction obtained from (a) by the unitary operator pumping $p+ip$ defect to the bottom and $p-ip$ defect to the top. Crucially, the top boundary, denoted as $f$ condensed*, though still have fermion condensed, differs from the usual $f$ condensed boundary by a $p-ip$ defect; (c) An auxiliary construction obtained from (b) by removing the $p-ip$ defect though modifying the Hamiltonian near the top boundary; (d) A sandwich that shares the same gapped bulk ground state, top and bottom gapped boundary state with (c), thus the Hamiltonians for (c) and (d) are related by a quasi-2D FDqLU. $S$ and $S'$ represent the bulk commuting Hamiltonian terms symbolically. And $S'$ is mapped from $S$ by the conjugation of the ``gauged p+ip defect'' pump unitary.}
    \label{fig:sandwiches_p+ip_constructions}
\end{figure}

\paragraph{Two ``sandwiches''} The generic approach to obtain a sandwich with a twisted bottom boundary using pump unitaries outlined in the beginning of Section 
\ref{sec:Invert_symTFT} does not apply in the $p+ip$ example. Since it would imply a locality preserving unitary between a trivial superconductor and a $p+ip$ superconductor, which is believed not possible. 
Let us provide a construction for the two ``sandwiches'' that only establishes a bijection between LEEs above the trivial and $p+ip$ superconductor ground state, while the two ground states are not related by a locality preserving unitary. The two sandwiches, shown as (a) and (d) in Figure~\ref{fig:sandwiches_p+ip_constructions}, are related through two auxiliary constructions in between. Let us explain the constructions in Figure~\ref{fig:sandwiches_p+ip_constructions}, and especially the relation between (a) and (d). 
\begin{enumerate}
    \item Sandwich (a) is to start with the bulk ground state of fermionic Toric code, a top gapped boundary where fermions are condensed together with physical fermions, and a bottom smooth boundary where flux loop are condensed (see Section \ref{subsubsec:fTC}). This construction after dimension reduction becomes a $2D$ gapped state with fermion parity symmetry, in other words, a trivial superconducting state.
    \item  Next, in the hope to prepare another sandwich differing from (a) by a bottom boundary with only relative difference, we can apply the unitary that pumps ``gauged $p+ip$ defect'' (see Section \ref{subsubsubsec:2toriccodes}) to (a), and thus obtain the sandwich (b). Nevertheless, crucially different from the previous SPT example, in this example, the top boundary fermion condensed boundary is not exactly invariant under the pump unitary. In other words, the fermion condensed boundary does not absorb the $p-ip$ defect. The new boundary after the unitary, still has fermion condensed on it, and is thus denoted as f condensed* boundary. Since both the defect and its inverse are present, sandwich (b) is equivalent to a trivial $2D$ superconductor after dimension reduction. 
    \item Now we remove the $p-ip$ defect near the top boundary by directly modify the Hamiltonian \emph{only} near the boundary. This leads to sandwich (c). This step is always possible. For example, we can stack an auxiliary layer of $p+ip$ Hamiltonian on top, and let the fermion couple with the bulk gauge field, such a layer will cancel out the $p-ip$ defect. Sandwich (c) is equivalent to a $p+ip$ superconductor after dimension reduction. 
    \item There exists further a quasi-2D FDqLU that maps the sandwich (c) to another sandwich (d) whose gapped ground state wavefunction is the same as (c), but the bulk Hamiltonian terms (symbolized by $S$ in the Figure) are the same as in the sandwich (a). 
\end{enumerate}
In summary, we obtain Sandwich (a) and (d) that have relatively different bottom boundaries, although the two sandwiches cannot be related by a 2D FDqLU.

We also illustrate the two sandwiches pictorially in Figure~\ref{fig:sandwich_p+ip}, where fermions on the top boundaries are condensed after bounded with physical fermions. An explicit Hamiltonian that describes the topological bulk phase and the fermion condensed phase is given in Appendix \ref{app:fTC3d}. The Hamiltonian where fermions are only condensed on the top boundary can be constructed similarly as in the $2d$ example shown in Figure~\ref{fig:ftop_stabilizers}.
The bottom boundaries of the two sandwiches, as they differ by an invertible defect, also only differ relatively. 

\begin{figure}[ht]
    \centering
    \begin{subfigure}[b]{0.95\textwidth}
   \includegraphics[width=1\linewidth]{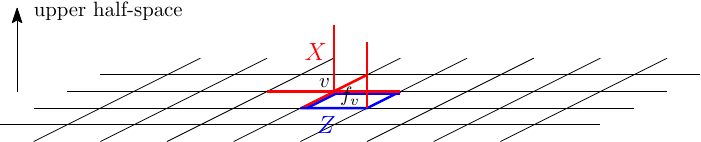}
   \caption{}
   \label{fig:fTC_half-space_stabilizer} 
    \end{subfigure}
    \begin{subfigure}[b]{0.95\textwidth}
   \includegraphics[width=1\linewidth]{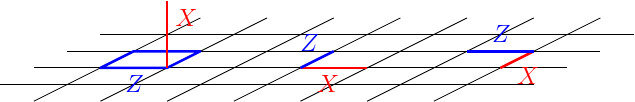}
   \caption{}
   \label{fig:fTC_bottom} 
    \end{subfigure}
    \caption{(a) A bulk stabilizer $A_vG_{f_v}$ fully supported in upper half-space, while individually neither $A_v$ or $G_{f(v)}$ is a bulk stabilizer.
    (b) A choice of generators of the operator algebra, red edge represents Pauli-$X$ and blue edge represents Pauli-$Z$. The terms are derived from combinations of truncated bulk terms $A_v$ and $G_f$ given in (\ref{eq:fTC}) near the bottom boundary.}
    \label{fig:ftc_bottom_operators}
\end{figure}

\paragraph{Dimension reduction, $\mathbb{Z}_2^f$ symmetry and boundary operator algebra} Sandwiches with fermionic Toric code ground state as the bulk state and fermion condensed top boundary are equivalent to quasi-2D systems with a global fermion parity symmetry. More explicitly, we can identify the operator mapping from a ``sandwich'' to the fermionic system with one fermion per face on a square lattice. First, to determine the operator algebra near the bottom boundary, we consider for example in a cubic lattice on upper half space that is terminated with a ``smooth'' cut, such that qubits on the bottom layer are located on edges of a $2d$ square lattice. The bulk state is topological and is the ground state of femionic Toric code. More precisely, the bulk state is stabilized by all combinations of $A_v$, $G_{f\in xy}$, $G_{f\in yz}$ and $G_{f\in zx}$ given in (\ref{eq:fTC}), provided they are fully supported within the upper half-space. For example, the term in Figure~\ref{fig:fTC_half-space_stabilizer} near the bottom layer is a bulk stabilizer because it is the product $A_vG_{f_v}$ which is fully supported within the upper half-space, though neither $A_v$ nor $G_{f_v}$ composing it is fully supported within the upper half-space. 

Next, we consider the algebra of local operators that commute with \emph{all} bulk stabilizers, which turn out to be all supported near the bottom boundary. We show a choice of generators of the algebra in Figure~\ref{fig:fTC_bottom}. Hamiltonians this algebra generates describe $\mathbb{Z}_2$ gauge theories with fermionic one-form symmetry \cite{chen2018exact}, with generators of the algebra shown in Figure~\ref{fig:fTC_half-space_stabilizer} at each vertex. The operators in Figure~\ref{fig:fTC_bottom} can be mapped to fermion bilinears in a $2d$ fermionic system. In other words, the operator algebra is isomorphic to the even fermionic algebra in $2d$. \cite{chen2018exact} The Hamiltonian terms are those illustrated in Figure~\ref{fig:fTC_bottom} weighted by different coupling constants.  In particular, with carefully chosen coupling constants, the Hamiltonians are mapped to those describing trivial superconductor and $p+ip$ superconductor. The latter cannot be a commuting projector model.  

The generator for the total fermion parity symmetry $\mathbb{Z}_2^f$ in two dimensions is mapped from the topological surface operator parallel to the boundaries, which represents the world history of a flux loop in fermionic toric code model. For example, it can be given by $\prod_{v\in \text{bottom layer}} X_{e(v)}$, where $e(v)$ is the vertical upper edge coming from $v$, and commutes with any terms in the operator algebra shown in Figure~\ref{fig:fTC_bottom}.

\paragraph{Bijection between LEEs of two sandwiches} With the constructions described in Figure~\ref{fig:sandwiches_p+ip_constructions}, we can obtain the bijections between LEEs in 2D trivial and topological superconductor, or equivalently in Sandwich (a) and (d). We can start with LEEs near the bottom boundary in Sandwich (b) and (c), since the bulk and bottom boundary Hamiltonian terms in (b) and (c) are exactly the same, operators creating LEEs are also identical. Then, LEEs near the bottom boundary in (a) are related to those in (b) via the conjugation by the ``gauged p+ip defect'' pump unitary (see Section \ref{subsubsec:fTC}); LEEs near the bottom boundary in (d) are related to those in (c) via the conjugation by the FDqLU that maps sandwich (c) to (d). Treating the sandwiches as quasi-2D system, LEEs near the bottom boundary reduces to LEEs respecting total fermion parity symmetry in the quasi-2D system. We thus establishes the bijections between symmetric LEEs in the trivial and $p+ip$ superconductor. Yet since sandwich (b) and (c) are not related by any 2D FDqLU, the bijection does not imply any locality preserving unitary between trivial and $p+ip$ superconductor.

\section{Discussion}
\label{sec:discussion}

In this paper, we study the low-entanglement excitations of invertible phases. Low-entanglement excitations of a gapped phase are expected to form a higher-category structure. For the trivial phase, it is the same as the $d$-category formed by gapped phases in $d$-dimension. In this paper, we show that for invertible phases, the same conclusion can be reached. We establish this result in several different ways. For SPTs within the group cohomology classification, different phases can be mapped to each other through a symmetric QCA, which can be used to prove the equivalence of their LEE classification. For chiral phases like the $p+ip$ superconductor, we used a `pumping' process in one higher dimension to argue for the equivalence of their LEE classification to their nonchiral counterparts. Moreover, we put the invertible phases into the Topological Holography framework and establish their equivalence through the equivalence of different boundary conditions of a topological theory in one higher dimension.

The result obtained in this paper extends our understanding of the non-fractional nature of invertible phases to excitations of higher dimensions. It is well understood that in invertible phases, point excitations cannot be fractionalized, and hence there are no anyon excitations or fractional charge excitations. This understanding was used, for example, to determine the possible Hall conductance in bosonic $U(1)$ symmetry-protected topological phases\cite{Senthil2013}. The result in this paper is a higher-dimensional version of this, saying that in invertible phases, not only point excitations, but also $d$-dimensional LEEs with $d>0$, have to be `non-fractional' in the sense that they are like $d$-dimensional gapped phases. This can lead to various interesting conclusions. For example, we can conclude that the flux line defect in the $2+1$D $p+ip$ superconductor takes the form of either a trivial or a nontrivial $1+1$D superconducting chain. Therefore, the endpoint of the flux line defect -- the $\pi$ flux -- can be a Majorana zero mode but not more fractionalized. 

\section*{Acknowledgement}
We are grateful for the inspiring discussions with Chong Wang. W.J. is grateful for the hospitality and partial support of the Kavli Institute for Theoretical Physics, where part of this work was performed, under NSF Grant PHY-2309135. X.C. is supported by the Walter Burke Institute for Theoretical Physics at Caltech, the Simons Investigator Award [award ID 828078] and the Institute for Quantum Information and Matter at Caltech. M.L. is supported in part by a Simons Investigator Award. The authors acknowledge support from the Simons Collaboration on Ultra-Quantum Matter [grant numbers 651438 (W.J. and X.C.), 651442 (M.L.), and 651440 (D.T.S)].

\appendix
\section{Fermion condensation}
The purpose of this appendix is to describe the fermion-condensed boundary condition for a topological order with emergent fermions using lattice models. Let us start by describing how to condense fermions in the bulk and begin with the example of toric code model in two dimensions. 

In the $\mathbb{Z}_2$ gauge theory, a fermionic quasiparticle can form a Cooper pair with a physical fermion and thus condense.\cite{aasen2019fermion} Let us describe a lattice model for the phase transition from the deconfined phase of the $\mathbb{Z}_2$ gauge theory to the fermion condensed phase. 

\begin{figure}[ht]
    \centering \includegraphics{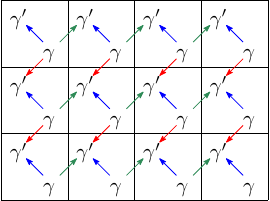}
    \caption{On each edge, there is a qubit. On each face, there is one complex fermions, composed of two Majorana fermions $\gamma,\gamma'$. The arrows specify the order to form a complex fermion and follow a Kasteleyn orientation \cite{cimasoni2007dimers,tarantino2016discrete,chen2023equivalence}, so that the repairing is well-defined.}
    \label{fig:latticeofqubitsandfermions}
\end{figure}

Our model is defined on a two-dimensional square lattice. As shown in Figure~\ref{fig:latticeofqubitsandfermions}, on each edge there is an qubit, representing the $\mathbb{Z}_2$ gauge field, and on each face, there is a complex fermion $c_f$ composed of two Majorana fermions $\gamma_f,\gamma'_f$, following the convension $\gamma_f=c_f+c_f^\dagger, \ii\gamma'_f=c_f-c_f^\dagger$. We assign arrows between majoranas to indicate the sign convention for the repairing of the majorana fermions into complex fermions and the hopping terms of majorana fermions, see Eq.~(\ref{eq:pairing}). \footnote{This assignment of arrows is also known as a Kasteleyn orientation. A Kasteleyn orientation is a lattice analogue of the spin structure. Specifically, it is an orientation on a planar graph for which any face has an odd number of clockwise-oriented edges.}\cite{cimasoni2007dimers,tarantino2016discrete}

The model has a hard gauge constrain for each face,
\begin{align}
    G_f\equiv A_{v(f)}B_f=1,~~A_v=\vcenter{\hbox{\includegraphics[scale=1.0]{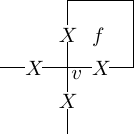}}},~~B_f=\vcenter{\hbox{\includegraphics[scale=1.0]{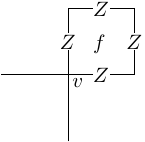}}},\label{eq:gaugeconstraint}
\end{align}
where $v(f)$ is the vertex on the left bottom corner of the face $f$. The Hamiltonian is the following,
\begin{align}
    H=-\sum_f B_f P_f^\gamma -h\sum_{f}B_f-t\sum_eU_eS_e 
    \label{eq:hamiltonian}
\end{align}
where 
\begin{align}
    P_f^\gamma=-\ii \gamma_f\gamma_f',~~S_e=\ii \gamma_{L(e)}\gamma'_{R(e)}, \label{eq:pairing}
\end{align} in which the arrow on the edge $e$ is pointing from $\gamma_{L(e)}$ to $\gamma_{R(e)}$. And
\begin{align}
U_e=\vcenter{\hbox{\includegraphics[scale=1.0]{fig/xzv.pdf}}}~~\text{or}~~\vcenter{\hbox{\includegraphics[scale=1.0]{fig/xzh.pdf}}}.
\label{eq:Ue1}
\end{align}
In the Hamiltonian, the term $U_eS_e$ imposes that the fermionic quasiparticle and the physical fermion hop in pairs across each edge. The term $B_fP_f^\gamma$ bonds the $\mathbb{Z}_2$ flux with a (physical) complex fermion on each face. It commutes with the second and third terms in the Hamiltonian, while the latter two anti-commute whenever $e\in \partial f$.

The Hamiltonian has two exactly solvable limits.
\begin{itemize}
    \item The deconfined phase, when $t=0,h>0$. The fermions are in the trivial insulating phase, and the qubits form a ground state of the usual toric code model. The ground state satisfies that
    \begin{align}
        A_v=1,~~B_f=1,~~P_f^\gamma=1.
    \end{align}
    On a square lattice with periodic boundary conditions, there are four degenerate ground states.
    \item The fermion condensed phase, when $h=0,t>0$. The ground states satisfy that
    \begin{align}
        B_fP_f^\gamma=1,~~U_eS_e=1.\label{eq:fcondensation}
    \end{align}
    And there are in total two degenerate ground states, as we show below. 
\end{itemize}

\subsection{Fermion condensed phase} The most transparent way to understand the fermion condensed phase is to first fermionize the $\mathbb{Z}_2$ gauge theory. The fermionization \cite{chen2018exact} maps the $\mathbb{Z}_2$ gauge theory with the gauge constraint (\ref{eq:gaugeconstraint}) to a fermion model on the square lattice, where on each face there is a complex fermion composed of two Majoranas $\chi,\chi'$. The face term $B_f$ and the hopping term $U_e$ of fermionic quasiparticles are mapped to fermion bilinears that preserve the same algebra (the even fermionic algebra \cite{chen2018exact}).
\begin{align}
    B_f\rightarrow P_f^\chi\equiv -\ii \chi_f\chi_f',~~~~U_e\rightarrow \ii\chi_{L(e)}\chi'_{R(e)}.
\end{align}

\begin{figure}
    \centering
    \includegraphics{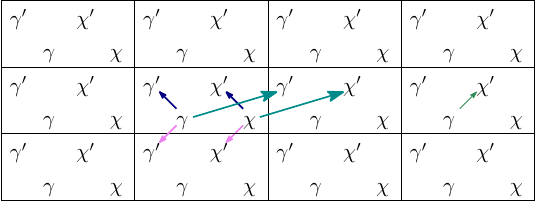}
    \caption{Four-Majorana-interacting terms. Each term represents that a pair of $\gamma$ and $\chi$ in one face hops to a neighboring pair of $\gamma'$ and $\chi'$, indicated by a pair of arrow of the same color. And there are three neighboring choices.}
    \label{fig:majoranainteraction}
\end{figure}

After fermionization, we obtain a model defined on a square lattice, where there are four Majorana fermions on each face, as shown in Figure~\ref{fig:majoranainteraction}. 
It follows that the solution (\ref{eq:fcondensation}) becomes
\begin{align}
    \left(-\ii \chi_f\chi_f'\right)\left(-\ii \gamma_f\gamma_f'\right)=1,~~\left(\ii\chi_{L(e)}\chi'_{R(e)}\right)\left(\ii\gamma_{L(e)}\gamma'_{R(e)}\right)=1.
    \label{eq:majoranainteraction}
\end{align}
We could illustrate these interacting terms pictorially, as shown in Figure~\ref{fig:majoranainteraction}. We see that each pair of majorana fermions $\gamma_f,\chi_f$ in a single face, appears in three four-Majorana-interacting terms. In each term, the pair of $\gamma$,$\chi$ hops to a neighboring pair of $\gamma'$,$\chi'$.

Thus, to satisfy all equations in (\ref{eq:majoranainteraction}), there are two solutions, 
\begin{align}
    \ii\chi_f\gamma_f=\ii\gamma'_f\chi_f'=r,~ \forall f;~~r\in\{1,-1\}.
    \label{eq:solution_fermion_condensed}
\end{align}

In terms of complex fermions, following the convention $\chi=a+a^\dagger, \ii \chi'=a-a^\dagger$, $\gamma=c+c^\dagger, \ii \gamma'=c-c^\dagger$, one can see that the two flavors of fermions are in a trivial superconducting phase. In particular, the fermion parity symmetry $P^\chi\equiv \prod_fP_f^\chi$ is spontaneously broken. This means, in the original model (\ref{eq:hamiltonian}), $\prod_fB_f$ is not conserved in the ground state.

\begin{figure}
    \centering
    \includegraphics{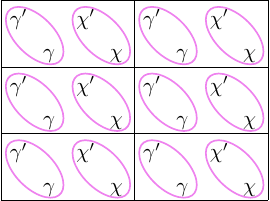}
    \hspace{50pt}
    \includegraphics{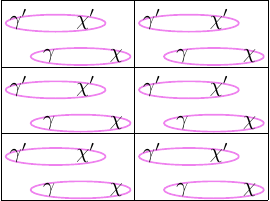}
    \caption{Two phases after fermionization. Left: fermionization of the deconfined phase with a trivial fermion insulator. The fermion parity symmetry of each flavor, $P^\gamma$ and $P^\chi$ are conserved. Right: fermion condensed phase. $P^\gamma$ and $P^\chi$ are broken; the total fermion parity $P^\gamma P^\chi$ is conserved. Here, $P^\gamma\equiv \prod_fP^\gamma_f$ and similarly for $P^\chi$. Every pair of majoranas in a circle represents a fermion bilinear term that acquires a classical value in the ground state.}
    \label{fig:enter-label}
\end{figure}

The fermion condensed phase with the solution (\ref{eq:solution_fermion_condensed}) preserves the total fermion parity, \begin{align}
    [\ii\chi_f\gamma_f,P]=[\ii\gamma_f'\chi_f',P]=0,~~~~P=\left(\prod_fP_f^\gamma\right)\left(\prod_fP_f^\chi\right).
\end{align}
In particular, when the boundary conditions are periodic for both directions, the total fermion parity is even, $P=1$.

Of course, since the fermion bilinears in the solution (\ref{eq:solution_fermion_condensed}) violate the single flavor fermion parity symmetry $P^\chi$, they cannot be mapped to a local operator in the original model using bosonization. Thus, it is not obvious how to represent the solutions after bosonization (or rather, to gauge the fermion parity symmetry $P^\chi$). Nevertheless, the degeneracy of ground states in the fixed point model after bosonization (see Eq.~(\ref{eq:fcondensation})) is still $2$, which we can compute independently by counting number of degrees of freedom as well as number of independent stabilizers. 

\subsection{$\mathbb{Z}_2$ gauge theory with a fermion-condensed boundary}\label{app:fcondensedbdry}
Now we would like to desribe the toric code model with a fermion condensed boundary, in the presence of physical fermions.

\begin{figure}[ht]
    \centering
        \begin{subfigure}[b]{0.8\textwidth}
    \centering
   \includegraphics[scale=0.8]{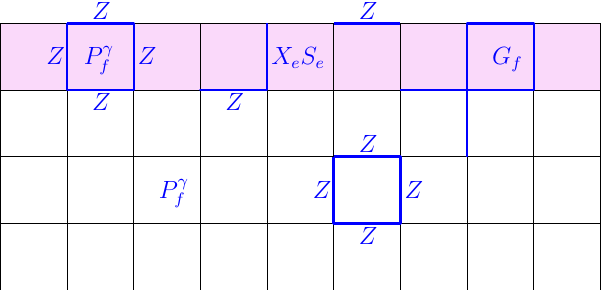}
   \caption{}
\end{subfigure}
\begin{subfigure}[b]{0.8\textwidth}
\centering
   \includegraphics[scale=0.8]{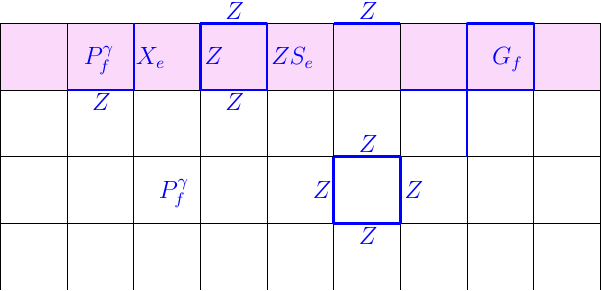}
   \caption{}
\end{subfigure}
    \caption{Stabilizers in the $\mathbb{Z}_2$ gauge theory with a fermion-condensed boundary. (a) untwisted $f$-condensed boundary. (b) Twisted $f$-condensed boundary. The two sets are related by the 2d unitary (\ref{eq:boundary_majorana_chain_pump}). This way of applying the unitary could be more natural to understand in the fermionized model. In that case, the pump operator translates all $\chi$-majoranas on the top row by one site, and then swaps the majorana pair $\chi$ and $\chi'$ in each face in a single step.}
    \label{fig:ftop_stabilizers}
\end{figure}

Consider the model described above but with a top boundary. Still the model satisfies the gauge constraint $G_f=1$, for all $f$. The Hamiltonian is as follows,
\begin{align}
    H&=H^{\text{bulk}}+H^{\text{bdry}},\nonumber\\
    H^{\text{bulk}}&=-\sum_{f\not\in\, \text{top row}}\left(B_fP_f^\gamma+ B_f\right),\nonumber\\
    H_f^{\text{bdry}}&=-\sum_{f\in\,\text{top row}}\left(B_fP_f^\gamma +U_{E(f)}S_{E(f)}\right)-\sum_{\text{top~}e}Z_e, \label{eq:fboundary}
\end{align}
where $E(f)$ represents the east edge of the face $f$. We illustrate these Hamiltonian terms and a gauge constraint term in Figure~\ref{fig:ftop_stabilizers} (a). 

\begin{figure}[ht]
    \centering   
    \begin{subfigure}[b]{0.8\textwidth}
    \centering
   \includegraphics{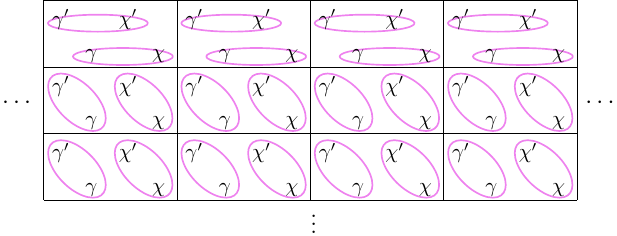}
   \caption{}
   \label{fig:ftopboundary2-1} 
\end{subfigure}
\begin{subfigure}[b]{0.8\textwidth}
\centering
   \includegraphics{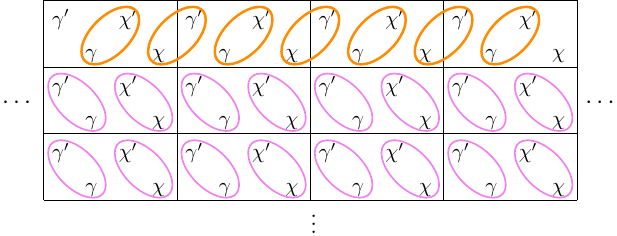}
   \caption{}
   \label{fig:ftopboundary2-2}
\end{subfigure}
    \caption{The ground state of the fermionization of $\mathbb{Z}_2$ gauge theory with a fermion-condensed boundary. Any fermion bilinear within a circle acquires a classical value when acting on the ground state. (a) Untwisted $f$-condensed boundary given in Eq.~(\ref{eq:fboundary}). (b) Twisted $f$-condensed boundary given in Eq.~(\ref{eq:twisted_fboundary}).}
    \label{fig:ftopboundary2}
\end{figure}

The model has two ground states. Besides the gauge constraint, both states satisfy that $B_f=P_f^\gamma=1$, for $f\not\in\,\text{top row}$, implying the toric code topological order in the bulk. The two states differ only near the boundary, which is easier to see after fermionization:
\begin{align}
    \ii \chi_f\gamma_f=\ii \gamma_f'\chi_f'=r_{\text{bdry}},~\forall f\in \,\text{top row};~~ r_{\text{bdry}}\in \{1,-1\}.
\end{align}
We illustrate these fermion bilinears that become classical values on the ground state in Figure~\ref{fig:ftopboundary2} (a). 

\begin{figure}[ht]
    \centering
    \includegraphics[scale=0.8]{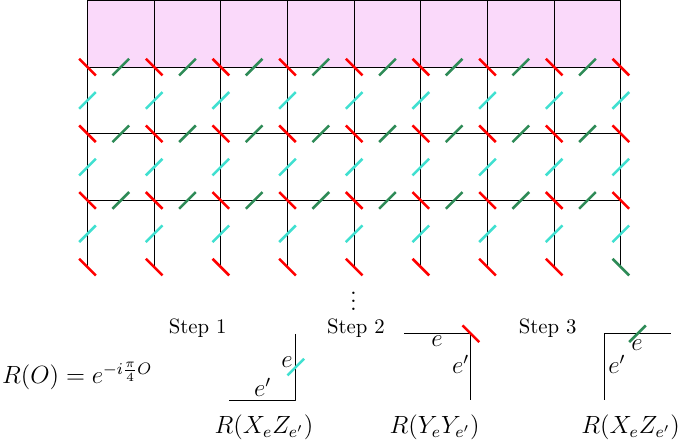}
    \caption{The unitary $U=V_{1d}U_{\text{pump}}$ in Eq.~(\ref{eq:boundary_majorana_chain_pump}) that pushes a gauged Kitaev chain defect to the boundary. $U_{\text{pump}}$ has three steps. In each step, one applies $\prod_{(ee')}R(O_{ee'})$ on pairs of edges $(ee')$ indicated by the color. $V_{1d}=\prod_{\text{violet }f}R(B_f)$.}
    \label{fig:boundary_majorana_chain_pump}
\end{figure}

\paragraph{$f$-condensed boundary stacked with gauged Kitaev chain defect} Now we twist the $f$-condensed boundary by the gauged Kitaev chain defect. The twisted boundary Hamiltonian can be obtained by conjugating $H^{\text{bdry}}$ in Eq.~(\ref{eq:fboundary}) by the unitary that pumps an gauged Kitaev chain/$e$-$m$ exchange defect to the top boundary. We choose to further apply a 1d FDLU near the top boundary to further simplify the boundary Hamiltonian. In the end, up to gauge constraint terms $G_f$, the twisted Hamiltonian is
\begin{align}
    U&H^{\text{bdry}}_f U^{-1} \simeq {H^{\text{bdry}}_f}',\nonumber\\
    {H^{\text{bdry}}_f}'=-\sum_{f\in\,\text{top row}}&\left(U_{E(f)}P_f^\gamma +B_{f+\hat{x}}S_{E(f)}\right)-\sum_{\text{top~}e}Z_e,
    \label{eq:twisted_fboundary}
\end{align}
where $E(f)$ is the edge on the east of the face $f$, and $f+\hat{x}$ is the neighbouring plaquette of $f$ on its east. And as we illustrated in Figure~\ref{fig:boundary_majorana_chain_pump}, the unitary $U$ is a 2d FDLU followed by a 1d FDLU,
\begin{align}
U=V_{\text{1d}}U_{\text{pump}},~~V_{\text{1d}}=\prod_{f\in\text{ top row}}e^{-\ii\frac{\pi}{4}B_f},\label{eq:boundary_majorana_chain_pump}
\end{align}
where the $U_{\text{pump}}$ is the 2d FDLU that pumps a gauged Kitaev chain/$e-m$ exchange line defect, as given Eq.~(\ref{eq:em_exchange_pump}) and acts on all qubits \emph{but those on the top horizonal and vertible edges}. The 1d FDLU, in terms of fermions, swaps the majoranas $\chi_f$ and $\chi'_f$ locally in each face $f$. 

We compare the difference between the stabilizers in the untwisted and untwisted boundary, as shown in Figure~\ref{fig:ftop_stabilizers}. 

\paragraph{Comment} We obtain the twisted version of the $f$-condensed boundary (\ref{eq:twisted_fboundary}) by applying the 2d FDLU to an untwisted $f$-condensed boundary. However, one can see that in (\ref{eq:twisted_fboundary}), the Hamiltonian terms within the brackets are nothing but to condense fermionic pairs, just as those terms in (\ref{eq:fboundary}). Thus, they are just two microscopic choices to condense fermionic pairs, and lead to the same macroscopic boundary condition -- fermion condensed boundary condition. We leave it as a future question, whether there exists additional categorical data to distinguish the two choices, whose boundary Hamiltonians cannot be related by a 1d FDLU.

We also remark that there are alternative approaches to constructing fermion-condensed boundaries in the 2d toric code model, where physical fermions are introduced only near the top boundary. There, the same idea of condensing pairs of emergent and physical fermions is applied, which one can tell using the operator mapping near the boundary shown in Figure~\ref{fig:operator_reduction_fermion}.


\subsubsection{Hamiltonians and a sequential circuit for ``sandwich constructions''}\label{subsubsec:Kitaev_chain_sandwich}
\begin{figure}[ht]
    \centering   
    \begin{subfigure}[b]{0.8\textwidth}
    \centering
   \includegraphics[scale=0.8]{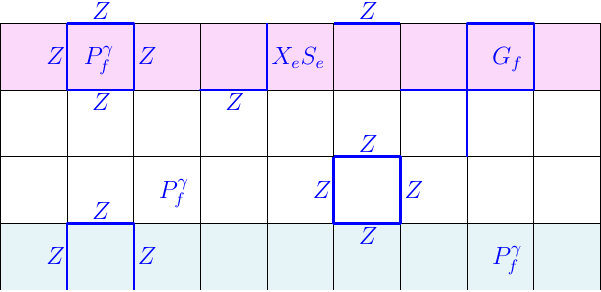}
   \caption{}
   \label{subfig:sandwich_f_tc_e}
\end{subfigure}
\begin{subfigure}[b]{0.8\textwidth}
\centering
   \includegraphics[scale=0.8]{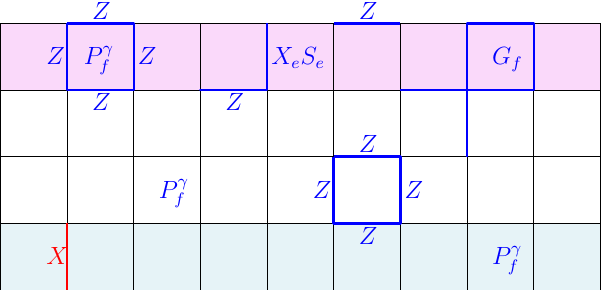}
   \caption{}
   \label{subfig:sandwich_f_tc_m}
\end{subfigure}
    \caption{Hamiltonian terms in the ``Sandwich construction'' from the toric code in 2D.}
    \label{fig:sandwich_f_tc_em)}
\end{figure}

We consider two sandwich constructions. The first one has the toric code model in the bulk with a $f$-condensed top boundary, and a $e$-condensed bottom boundary. We illustrate the terms  in the fixed-point Hamiltonian in Figure~\ref{subfig:sandwich_f_tc_e}. The full Hamiltonian is a summation of these terms and their translational counterparts, with an overall minus sign. The second ``sandwich'' has the same bulk and top boundary, but has $m$-anyon condensed on the bottom boundary. The terms in the fixed-point Hamiltonian, up to their translational counterparts and an overall minus sign, are given in Figure~\ref{subfig:sandwich_f_tc_m}. 

 The two Hamiltonians for the two sandwiches, are related by a sequential circiuit inserting a single $e-m$ exchange defect (given by Eq.~(\ref{eq:em_exchange_sequential_circuit}) in the bulk) near the bottom boundary. More precisely, for qubits on the edges along the bottom boundary, let us index those on the vertical (horizontal) edges using integers $i$ (half-integers $i+\frac{1}{2}$) with $i=1,2 \cdots, N$. So in Figure~\ref{subfig:sandwich_f_tc_e}, the three-body term on the bottom boundary is $Z_{i-1} Z_{i-\frac{1}{2}}Z_{i}$ and in Figure~\ref{subfig:sandwich_f_tc_m}, the single Pauli-$X$ term is $X_i$. The sequential circuit that swaps these two types of bottom boundary Hamiltonian terms is
\begin{align}
    U = R(Z_N Z_{N+\frac{1}{2}}Z_1) R(X_N)\cdots R(Z_{i}Z_{i+\frac{1}{2}}Z_{i+1})R(X_i)\cdots R(Z_1Z_{\frac{3}{2}}Z_2)R(X_1),
    \label{eq:boundary_circuit}
\end{align}
where $R(O)=e^{\ii \frac{\pi}{4}O}$. We obtain the circuit by generalizing the sequential circuit for Kramers-Wannier duality \cite{chen2024sequential} to the boundary of the $\mathbb{Z}_2$ gauge theory.

Within the subspace 
\begin{align}
    W_m(C^\vee_{\text{bdy}})=\prod_{i=1}^N X_i=1,~~W_e(C_{\text{bdy}})=\prod_{i=1}^N Z_{i+\frac{1}{2}}=1,
\end{align} 
any local operator is mapped to a local operator by the circuit (\ref{eq:boundary_circuit}). In particular, under the conjugation by $U$, 
\begin{align}
    Z_iZ_{i+\frac{1}{2}}Z_{i+1} &\rightarrow X_i,~\forall i, \\
    X_1 & \rightarrow W_m(C^\vee_{\text{bdy}})W_e(C_{\text{bdy}}), \\
    X_{i} &\rightarrow Z_{i-1}Z_{i-\frac{1}{2}}Z_i,~ i = 2,\cdots, N,
\end{align}
In other words, the circuit is a locality preserving unitary \emph{within} the subspace $W_m(C^\vee_{\text{bdy}})=1,\; W_e(C_{\text{bdy}})=1$. After dimension reduction, we consider the sandwich as a quasi-1d system with $\mathbb{Z}_2^f$ symmetry, the circuit becomes a locality preserving unitary within the fermion parity even sector. Indeed, through the inverse of Jordan-Wigner transformation, the circuit becomes the unitary that translates all majoranas by one.

\subsection{Fermion condensation in the fermionic Toric code in $3d$}\label{app:fTC3d}
Our lattice description to condense emergent fermion in a topological order can be generalized to higher dimensions. The simplest example is to consider fermionic toric code in three dimensions. 

On a cubic lattice in three dimensions, we have one qubit on each edge $e$, and two Majorana fermions on each vertex $v$. The model has a (fermionic) gauge constraint \cite{chen2021higher} for each face $f$ ,
\begin{align}
    G_f\equiv\vcenter{\hbox{\includegraphics[scale=1.0]{fig/fTC_dual_a.pdf}}}~\text{or}~~\vcenter{\hbox{\includegraphics[scale=1.0]{fig/fTC_dual_b.pdf}}} ~\text{or}~~ \vcenter{\hbox{\includegraphics[scale=1.0]{fig/fTC_dual_c.pdf}}}=1,
\end{align}
where on each red(blue) edge, there is a Pauli-$X$(Pauli-$Z$) operator.
The Hamiltonian is
\begin{align}
    H&=-\sum_v A_v P_v - h\sum_v A_v,-t\sum_e U_eS_e,\\
    A_v&=\prod_{e:v\in \partial e}Z_e,~~P_v=-\ii \gamma_v\gamma_v',\nonumber\\
    U_e& =\vcenter{\hbox{\includegraphics{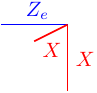}}}~\text{or}~~\vcenter{\hbox{\includegraphics{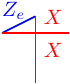}}}~\text{or}~~\vcenter{\hbox{\includegraphics{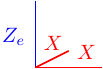}}},\nonumber\\
    S_e&=\ii \gamma_{v_L(e)}\gamma_{v_R(e)}'\nonumber.
\end{align}

Similar to the two-dimensional case, this model also has two exactly solvable limits. 
\begin{itemize}
    \item The deconfined phase, when $t=0,h>0$. The fermions are in the trivial insulating phase. The solution is
    \begin{align}
        G_f=1,~~A_v=1,~~P_v^\gamma=1.
    \end{align}
    
    \item The fermion condensed phase, when $h=0,t>0$. The solution is
    \begin{align}
        A_vP_v^\gamma=1,~~U_eS_e=1.\label{eq:3dfcondensation}
    \end{align}
\end{itemize}

\section{A sequential circuit generating a Kitaev chain defect line}
\begin{figure}[ht]
    \centering
    \includegraphics{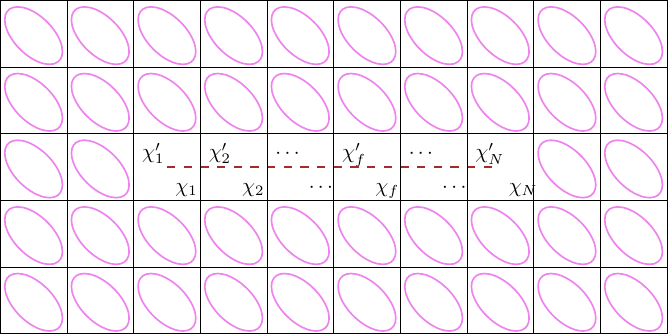}
    \caption{A defect line in a two-dimensional trivial fermion superconductor.}
    \label{fig:defect_line}
\end{figure}

In terms of Majorana fermion operators, the circuit is \cite{huang2015quantum,chen2024sequential}
\begin{align}
    U_F=e^{\frac{\pi}{4}\chi_N'\chi_1}\prod_{f=N-1}^1 e^{\frac{\pi}{4}\chi_{f+1}'\chi_{f+1}}e^{\frac{\pi}{4}\chi_{f}\chi_{f+1}'},
\end{align}
where $f$ run over the faces along the defect line, as shown in Figure~\ref{fig:defect_line}.

We would like to perform a {\it{2D bosonization}}\cite{chen2018exact} on this circuit, which preserves the total fermion parity symmetry in the two dimensional system. In particular,
\begin{align}
    -\ii \chi_f\chi_{f}'&\rightarrow B_f=\prod_{e\in \partial f}Z_e,\nonumber\\
    -\ii \chi_f\chi_{f+1}'&\rightarrow U_{e_R(f)}=X_{e_R(f)}Z_{e'}.
\end{align}
Thus, neglecting the first (non-local) term $e^{\frac{\pi}{4}\chi_N'\chi_1}$, the circuit becomes
\begin{align}
    U=\prod_{f=N-1}^1R(B_{f+1})R(U_{e_R(f)}),~~R(O)=e^{-\ii\frac{\pi}{4}O}.
    \label{eq:em_exchange_sequential_circuit}
\end{align}

\begin{figure}[H]
    \centering
    \includegraphics{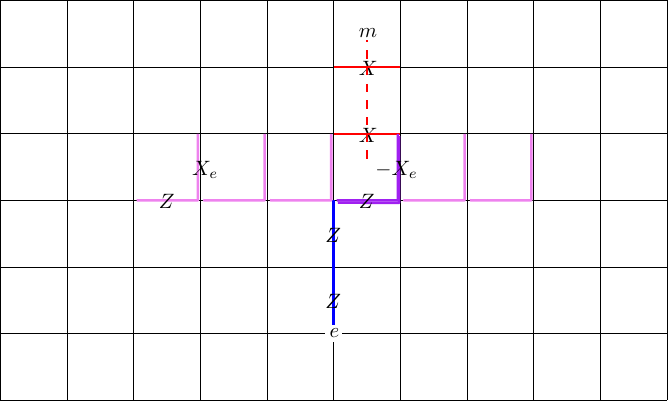}
    \caption[]{Stabilizers of the toric code in the presence of a $e-m$ exchange defect line. Along the line, there is a stabilizer $X_eZ$ defined on the violet edges for each plaquette. When the Wilson line creating $m$ anyons (the $\pi$ flux) crosses the defect line, the stabilizer at the crossing is changed to $-X_eZ$\footnotemark; and the $m$-type Wilson line crosses the defect line and becomes the $e$-type Wilson line. Such stabilizers on the defect line can be easily understood through the fermionization of the toric code, as shown in Figure~\ref{fig:kitaev_chain}.}
    \label{fig:em_defect_line}
\end{figure}

\paragraph{EM exchange defect line as an emergent symmetry}Along the $e$-$m$ exchange defect line, the stabilizer is changed, for any face $f$ along the defect line,
\begin{align}
    B_f=\prod_{e\in \partial f} Z_e\rightarrow U_{E(f)},
\end{align}
where $E(f)$ is the edge of the face $f$ on its east. $U_e$ is a ``short-string'' operator that hops a quasi-particle fermion across the edge $e$. Depending on the orientation of the edge $e$, it takes the form
\begin{align}
U_e=\vcenter{\hbox{\includegraphics[scale=1.0]{fig/xzv.pdf}}}~~\text{or}~~\vcenter{\hbox{\includegraphics[scale=1.0]{fig/xzh.pdf}}}.
\end{align}
We give an illustration of stabilizers on the defect line in Figure~\ref{fig:em_defect_line}.

\footnotetext{The stabilizers (one for each face) on the defect line are not to be confused with a Wilson line creating fermionic quasiparticle $\epsilon=e\times m$, which is a single string operator.}

\begin{figure}[H]
    \centering
    \includegraphics{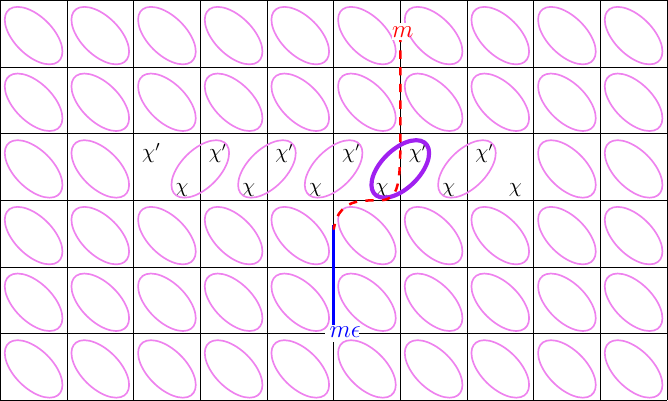}
    \caption{A Kitaev chain in the background of trivial fermion superconductor. Along the chain, the Majorana fermions neighboring an edge $e$ (violet circle across the edge) are paired as $\ii \chi_{L(e)}\chi_{R(e)}'$. When a $\pi$-flux (i.e., the fermion parity flux) goes across the Kitaev chain, the fermion bilinear term becomes $-\ii \chi_{L(e)}\chi_{R(e)}'$ (highlightened by the purple circle). A complex fermion mode is thus pumped to the $\pi$-flux line. \cite{PhysRevB.105.045106,rao2021theory} Each violet circle within a face $f$ represents $-\ii \chi_f\chi_f'$, whose ground state provides the trivial superconducting background.}
    \label{fig:kitaev_chain}
\end{figure}

\bibliography{references}
\end{document}